\documentclass[a4paper,11pt]{article}
\usepackage{amsmath,amssymb,color,comment}
\usepackage[normalem]{ulem}
\usepackage{slashed, tensor, bm, physics}
\usepackage{caption}
\usepackage{graphicx}
\usepackage{multirow}
\usepackage{float}
\usepackage[compat=1.1.0]{tikz-feynhand}
\usepackage{jheppub} 

\usepackage[T1]{fontenc} 
\usepackage{booktabs} 

\title{A particle on a ring or: how I learned to stop worrying and love $\theta$-vacua}
\author[a]{Mohammad Aghaie}
\author[a]{and Ryosuke Sato}

\affiliation[a\,]{Department of Physics, The University of Osaka, Toyonaka, Osaka 560-0043, Japan}

\emailAdd{aghaie@het.phys.sci.osaka-u.ac.jp}
\emailAdd{rsato@het.phys.sci.osaka-u.ac.jp}

\abstract{
Recently, Ai, Cruz, Garbrecht, and Tamarit (ACGT)~\cite{Ai:2020ptm, Ai:2024cnp, Ai:2025quf} claimed that the strong CP problem can be avoided by adopting a particular order of limits in the Euclidean path integral, in which the spacetime volume is taken to infinity before summing over all topological sectors. We critically examine this proposal using exactly solvable examples of one-dimensional quantum mechanics on a ring, namely the quantum rotor and the quantum pendulum. These systems provide fully controlled settings with known $\theta$-dependent spectra. We find that the ACGT procedure fails to reproduce the correct energy spectrum. Since the spectrum is a direct physical observable, this result demonstrates that the proposed order of limits cannot be justified and conclusions about CP conservation in QCD cannot be based on this prescription alone.}

\begin{document} 
\begin{flushright}
OU-HET-1296
\end{flushright}
\maketitle
\flushbottom

\section{Introduction}
Despite the fact that Quantum Chromodynamics (QCD) generically allows for CP violation, no such effects have been observed in hadronic phenomena, giving rise to the so-called strong CP problem~\cite{Callan:1976je, Jackiw:1976pf}. At the level of the QCD Lagrangian, CP violation can arise from two independent sources: the topological $\theta$ term, $(\theta g_s^2 / 32\pi^2)\, G^a_{\mu\nu} \tilde G^{a\mu\nu},$
and complex phases in the quark mass matrix. After accounting for chiral rotations of the quark fields, these contributions combine into a single physical CP-violating parameter, \(\bar{\theta} = \theta + \arg\det M_q\). Since no fundamental symmetry enforces $\bar{\theta}=0$, one would naturally expect CP-violating effects in strong interactions, such as a sizable neutron electric dipole moment. However, experimental bounds on the neutron EDM imply the extremely stringent constraint $|\bar{\theta}| \lesssim 10^{-10}$~\cite{Abel:2020pzs, Chupp:2017rkp}. The tension between the generic expectation of $\mathcal{O}(1)$ CP violation in QCD and its apparent absence in nature constitutes the strong CP problem.

The conventional wisdom is that the strong CP problem is an unavoidable consequence of resolving the $U(1)_A$ problem (see, e.g., ref.~\cite{Gabadadze:2002ff} for a review). Chiral symmetry breaking in QCD naively predicts a nonet of Nambu–Goldstone bosons whose mass squareds are proportional to the light quark masses. However, the mass of the ninth Nambu–Goldstone boson, the $\eta'$, is larger than the prediction of the chiral Lagrangian~\cite{Weinberg:1975ui}. This puzzle was resolved by the discovery of a topologically nontrivial gauge-field configuration which breaks the axial $U(1)_A$ symmetry~\cite{Belavin:1975fg, tHooft:1976rip, tHooft:1976snw}. Such a configuration necessarily introduces a phase parameter $\bar\theta$ in the definition of the vacuum~\cite{Callan:1976je, Jackiw:1976pf}, and a nonzero $\bar\theta$ generally violates CP symmetry. Indeed, Shifman, Vainshtein, and Zakharov~\cite{Shifman:1979if} showed that the large $\eta'$ mass implies a nonvanishing topological susceptibility $\chi$, which leads to physical CP-violating effects for nonzero $\bar\theta$. Based on this understanding, the neutron EDM has been studied using chiral perturbation theory~\cite{Crewther:1979pi, Pich:1991fq}, QCD sum rules~\cite{Pospelov:1999ha, Pospelov:1999mv, Pospelov:2005pr, Hisano:2012sc, Ema:2024vfn}, and lattice QCD~\cite{Dragos:2019oxn, Alexandrou:2020mds, Bhattacharya:2021lol, Liang:2023jfj, He:2023gwp, Liu:2024kqy}. Although these approaches involve $\mathcal{O}(1)$ uncertainties, they consistently predict $d_n \sim \bar{\theta} \times 10^{-13}~e~\mathrm{cm}$. In addition to the neutron EDM, a nonzero $\bar{\theta}$ induces an electron EDM~\cite{Flambaum:2019ejc, Choi:1990cn, Ghosh:2017uqq} and CP-violating decays such as $\eta \to \pi\pi$~\cite{Shifman:1979if, Cheng:1987gp}. Although neither has been observed experimentally~\cite{Roussy:2022cmp, KLOE-2:2020ydi}, they provide complementary, albeit weaker, constraints on $|\bar{\theta}|$ compared to the neutron EDM~\cite{Pospelov:2025vzj}. Furthermore, the relation between the $\eta'$ mass and the $\bar\theta$ dependence has been extensively discussed in large-$N_c$ QCD~\cite{Witten:1979vv, Veneziano:1979ec} and in supersymmetric gauge theories~\cite{Dine:2016sgq, Csaki:2023yas}.

Recently, Ai, Cruz, Garbrecht, and Tamarit (ACGT)~\cite{Ai:2020ptm, Ai:2024cnp, Ai:2025quf} have tried to challenge this foundation, arguing that the $\bar\theta$-dependence is an artifact of an incorrect order of limits in the Euclidean path integral.\footnote{
For other recent trials, see, e.g., refs.~\cite{Yamanaka:2022vdt, Yamanaka:2022bfj, Yamanaka:2024nzn, Nakamura:2021meh, Schierholz:2024var, Schierholz:2025tns, Strocchi:2024tis}.}
According to them, if the infinite spacetime volume limit ($T \to \infty$) is taken before summing over topological sectors (winding numbers), the $\bar\theta$-dependence vanishes from all physical observables, thereby eliminating the strong CP problem without new physics.
Also, they tried to justify this formulation by using canonical quantization \cite{Ai:2024vfa}.

This proposal has ignited a significant debate in the community and there have been a few responses from different perspectives. Benabou \emph{et al.}~\cite{Benabou:2025viy} critiqued the proposal from the perspective of effective field theory and the phenomenology of the $\eta'$ meson, while Khoze~\cite{Khoze:2025auv} demonstrated that, in ACGT procedure, the partition function in a large volume limit has an additional factor, which has the square root dependence on the volume. This prevents the extraction of the vacuum energy from the partition function. A particularly concrete test was provided by Albandea \emph{et al.}~\cite{Albandea:2024fui}, who used lattice simulations of a quantum-mechanical rotor to show that local observables retain their $\bar\theta$-dependence, and that a direct application of the ACGT prescription leads to unphysical results. In response, ACGT~\cite{Ai:2025quf} have refined their argument, distinguishing between global and local definitions of topological susceptibility to maintain consistency with observed phenomena.

Moreover, Bhattacharya~\cite{Bhattacharya:2025qsk} argued that in a gapped quantum field theory such as QCD, the effect of boundary conditions on local observables is exponentially suppressed with the distance from the boundary. 
Consequently, although ref.~\cite{Ai:2020ptm} argued that a boundary condition on the gauge field affects the calculation of the topological susceptibility, a boundary-condition-dependent term only arises as an exponentially suppressed term in a physical quantity,
explaining why open and zero-field boundary conditions are physically equivalent for locally defined observables in the infrared limit, despite being mathematically distinct. This work might be the most direct response to the shortcoming suggested by ACGT.

More recently, Gamboa \emph{et al.}~\cite{Gamboa:2025hxa} challenged the ACGT claim by stressing the distinction between local and global observables, arguing that the disappearance of $\theta$ from certain fermionic correlators does not signal CP conservation but rather reflects the insensitivity of local probes to global holonomy. In contrast, genuinely global quantities like topological susceptibility remain $\theta$-dependent. Also, Ringwald \cite{Ringwald:2026apz} argued that the vacuum energy of one-flavor QCD should be a real-analytic function of the quark mass due to the existence of a mass gap, a property that is incompatible with the chiral Lagrangian proposed by ACGT.

While the original arguments are formulated in four-dimensional gauge theory, their essential logic can be tested in simpler quantum-mechanical models where calculations are fully controllable. Two standard examples are particularly instructive: a free particle on a ring (the quantum rotor) and a particle on a ring subject to a periodic potential (the quantum pendulum). Both exhibit nontrivial topology and a $\theta$ dependence, but they emphasize complementary aspects of the problem and come with distinct limitations.

The quantum rotor is the simplest exactly solvable system with nontrivial topology: its full spectrum and propagator can be derived analytically, and the latter requires an explicit sum over winding numbers, making the $\theta$ dependence manifest. However, the absence of a potential means that the wavefunctions are completely delocalized, obscuring the notion of distinct “$n$-vacua.” By contrast, the quantum pendulum provides a more direct semiclassical analogy to gauge theory, with potential minima at $\phi=2\pi n$ representing classical vacua and instanton tunneling producing a $\theta$-dependent Bloch ground state. This picture is physically transparent but not exactly solvable, as analytic control typically relies on semiclassical or WKB approximations rather than a fully explicit path-integral construction.

Here, we revisit the issue in these fully controlled quantum-mechanical systems where all relevant observables can be computed analytically via two equivalent methods: canonical quantization and the Euclidean path integral. While the canonical formulation provides an unambiguous benchmark, the path-integral approach allows us to explicitly implement the ACGT prescription and directly examine its consequences for important quantities such as energy spectrum, propagator, partition function, and topological susceptibility. We show that taking the Euclidean time extent $\beta\to\infty$ prior to summing over topological sectors removes the contribution of nontrivial winding configurations, leading to physically inconsistent results. 
Importantly, the established $\theta$ dependence in the energy spectra of both systems is successfuly reproduced when the summation over all topological sectors is performed \textit{before} taking the $\beta\to\infty$ limit.

We also compute the topological susceptibility in both the canonical and Euclidean path-integral formulations and explicitly verify their equivalence when the relevant limits are taken in the physically consistent order. We show that
starting from the main definition of the topological susceptibility as the curvature of the vacuum energy density~\cite{Witten:1979vv, Veneziano:1979ec}, one can reproduce the established quantum mechanical results only when the contributions of all nontrivial topological sectors are included prior to taking the infinite-time limit. Otherwise, the susceptibility vanishes identically, in contradiction with exact canonical quantum mechanics and lattice results.

As a direct consequence, the order-of-limits proposed by ACGT leads to the phenomenological inconsistencies emphasized by Benabou \emph{et al.}~\cite{Benabou:2025viy}. In particular, it implies a vanishing topological susceptibility in pure Yang–Mills theory (and in full QCD), in contradiction with lattice results and established phenomenology. Moreover, in the large-$N_c$ limit the Witten–Veneziano relation directly ties the $\eta'$ mass to the Yang–Mills topological susceptibility, so a vanishing susceptibility is also incompatible with the observed $\eta'$ mass.

This paper is organized as follows. In section~\ref{sec:quantum-rotor}, we analyze the quantum rotor, reviewing its canonical quantization, propagator, and partition function, and show how the summation over topological sectors in the path-integral formulation leads to a $\theta$-dependent energy spectrum and a nonvanishing topological susceptibility. We also present a perturbative evaluation and clarify the role of the order of limits. In section~\ref{sec:quantum-pendulum}, we extend the discussion to the quantum pendulum and reexamine the computation of the topological susceptibility using the prescription proposed by ACGT, highlighting how their order-of-limits procedure alters the contribution of topological sectors. Finally, section~\ref{sec:conclusion1} contains a discussion of the implications of these results for $\theta$ dependence and the strong CP problem,
and present our conclusion in section \ref{sec:conclusion}.

\section{A ring without potential: quantum rotor}
\label{sec:quantum-rotor}
The quantum rotor is the simplest exactly solvable quantum-mechanical system whose compact, multiply connected configuration space gives rise to distinct winding sectors. Importantly, its dynamics is simple enough to be exactly solvable in both the canonical and path-integral formulations. Therefore, we use the quantum rotor as a toy model to investigate the conceptual issues raised by ACGT in the context of QCD, focusing in particular on the order of limits proposed in~\cite{Ai:2020ptm}, where the spacetime volume is taken to infinity before summing over all winding sectors.

To this end, we begin by solving the Schr\"odinger equation for this toy model in order to set the stage and fix our notation. We then compute the propagator using both approaches and demonstrate their equivalence. Finally, we use this framework to analyze the topological susceptibility and examine the consequences of taking the infinite-time limit prior to summing over topological sectors, thereby clarifying how different prescriptions affect the recovery of standard quantum-mechanical results.

Throughout this paper, we work in natural units and set $\hbar = 1$. The real-time and Euclidean time extents are denoted by $T$ and $\beta$, respectively.

\subsection{Canonical quantization and energy spectrum}
Let us consider a free particle of mass $m$ constrained to move on a ring of radius $R$. The Lagrangian for this system is
\begin{equation}
    L(\phi, \dot{\phi}) = \frac{1}{2} I \dot{\phi}^2 + \frac{\theta}{2\pi}\dot{\phi},
    \label{eq:gen_lag}
\end{equation}
where $I = m R^2$ is the moment of inertia and $\theta$ is a vacuum angle conjugate to the winding number. As is well known, the second term is a total derivative and therefore does not contribute to the classical equations of motion. Nevertheless, when the configuration space satisfies $\phi \sim \phi + 2\pi$, this term has nontrivial quantum-mechanical consequences.
As shown in the appendix \ref{sec:charged particle}, the effect of $\theta$ can be understood as Aharonov-Bohm effect on a charged particle.

In order to isolate the role of topology, we focus in this section on the simplest case with vanishing potential and, for the moment, ignore the explicit $\theta$-term in the Lagrangian. Using the canonical momentum $p = I \dot{\phi}$, the Hamiltonian of the system takes the form
\begin{equation}
    H = -\frac{1}{2I}\frac{\partial^2}{\partial \phi^2}.
    \label{eq:H_rotor}
\end{equation}
The time-independent Schr\"odinger equation (TISE) is therefore
\begin{equation}
    -\frac{1}{2I}\frac{\partial^2 \psi}{\partial \phi^2} = E \psi(\phi).
    \label{eq:schroedinger_eq_free}
\end{equation}
Since the angular coordinate $\phi$ is periodic, $\phi \sim \phi + 2\pi$, the most general boundary condition consistent with a single-valued probability density is
\begin{equation}
    |\psi(\phi + 2\pi)|^2 = |\psi(\phi)|^2
    \quad \Rightarrow \quad
    \psi(\phi + 2\pi) = e^{i\theta}\psi(\phi),
    \label{eq:twisted_bc}
\end{equation}
which is known as the twisted boundary condition. The parameter $\theta \in [0,2\pi)$ plays the role of a vacuum angle, analogous to the $\theta$-parameter in Yang--Mills theory.

Solving eq.~\eqref{eq:schroedinger_eq_free} subject to the boundary condition \eqref{eq:twisted_bc}, we obtain the normalized eigenfunctions
\begin{equation}
    \psi_n(\phi) = \frac{1}{\sqrt{2\pi}} e^{i k_n \phi},
    \qquad
    k_n = n + \frac{\theta}{2\pi},
    \qquad
    n \in \mathbb{Z},
    \label{eq:eigenfunctions}
\end{equation}
with corresponding energy eigenvalues
\begin{equation}
    E_n(\theta) = \frac{  k_n^2}{2I}
    = \frac{1}{2I}\left(n + \frac{\theta}{2\pi}\right)^2.
    \label{eq:spectrum_free}
\end{equation}

An equivalent description is obtained if we keep the $\theta$-term in the Lagrangian, $\mathcal{L}\supset \frac{\theta}{2\pi}\dot{\phi}$. In this case, the canonical momentum is shifted by $p \to p  +\theta/2\pi$, and the wavefunctions are strictly periodic, $\psi(\phi+2\pi)=\psi(\phi)$, while the Hamiltonian is modified by a corresponding momentum shift. The resulting energy spectrum is identical to \eqref{eq:spectrum_free}, demonstrating the equivalence between twisted boundary conditions and the inclusion of a topological $\theta$-term at the canonical level.

\subsection{Propagator}
\label{sec:propagatorsec}
The propagator of the quantum rotor can be computed exactly in two complementary ways. Canonical quantization, via the Schrödinger equation, yields the well-known energy spectrum and its explicit $\theta$ dependence. Independently, the path-integral formulation expresses the propagator as a sum over winding sectors with an explicit $\theta$ term, making the role of topology and the order of limits manifest. This allows a direct test of the ACGT order-of-limits prescription against the exact canonical result. In the following, we present both approaches.

\subsubsection*{Method A: Canonical Quantization}

The propagator is defined as
\begin{align}
    K(\phi_f,t_f;\phi_i,t_i)
    &= \langle \phi_f | e^{-iH(t_f - t_i) } | \phi_i \rangle \notag\\
    &= \sum_n \langle \phi_f | n \rangle \langle n | \phi_i \rangle
    e^{-iE_n (t_f - t_i) }.
\end{align}
Using the eigenfunctions $\psi_n(\phi) = e^{in\phi}/\sqrt{2\pi}$, this expression becomes 
\begin{align}
    K(\phi_f,t_f;\phi_i,t_i)
    &= \frac{1}{2\pi} \sum_{n=-\infty}^{\infty}
    e^{i n (\phi_i - \phi_f)}
    e^{-iE_n T },
    \qquad T = t_f - t_i \notag\\
    &= \frac{1}{2\pi} \sum_{n=-\infty}^{\infty}
    \exp\!\left[
        i n (\phi_i - \phi_f)
        - \frac{i T}{2I}\left(n + \frac{\theta}{2\pi}\right)^2
    \right] \nonumber\\
    &= \frac{1}{2\pi} \exp\left( -\frac{i\theta^2 T}{8\pi^2 I} \right) \vartheta\left(-\frac{\phi_f-\phi_i}{2\pi} -\frac{ \theta T}{4I\pi^2};~ -\frac{ T}{2\pi I} \right),
    \label{eq:propagator_Sch}
\end{align}
where $\vartheta(z;\tau)$ is Jacobi theta function defined as
$\vartheta(z;\tau) = \sum_{n=-\infty}^\infty \exp\left( i \pi \tau n^2 + 2\pi i n z \right)$.
See also refs.~\cite{LeinoThesis, Benabou:2025viy}.

\subsubsection*{Method B: Path Integral}

We now compute the same propagator using the path integral formulation. Setting $t_i = 0$ and $t_f = T$ for convenience, the propagator can be written as
\begin{equation}
    K(\phi_f,T;\phi_i,0)
    =
    \int_{\phi(0)=\phi_i}^{\phi(T)=\phi_f}
    \mathcal{D}\phi\;
    e^{i S[\phi]}.
\end{equation}
The action, including a topological $\theta$-term, is
\begin{equation}
    S_{\theta}[\phi] =
    \int_0^T dt\,
    \left[
        \frac{I}{2}\dot{\phi}^2
        + \frac{\theta}{2\pi}\dot{\phi}
    \right].
\end{equation}
Then, the path integral becomes
\begin{align}
    K(\phi_f,T;\phi_i,0)
    &=
    \sum_{k=-\infty}^{\infty}
    \int_{\phi(0)=\phi_i}^{\phi(T)=\phi_f + 2\pi k}
    \mathcal{D}\phi\;
    \exp\!\left[
        i S_0[\phi]
        + \frac{i\theta}{2\pi}(\phi_f - \phi_i + 2\pi k)
    \right] \notag\\
    &=
    \exp\!\left[\frac{i\theta}{2\pi}(\phi_f - \phi_i)\right]
    \sum_{k=-\infty}^{\infty}
    e^{ik\theta}
    \int_{\phi(0)=\phi_i}^{\phi(T)=\phi_f + 2\pi k}
    \mathcal{D}\phi\;
    \exp\!\left[i S_0[\phi]\right].
    \label{eq:propagator_circ}
\end{align}
The remaining path integral corresponds to a free particle on the real line,
\begin{equation}
    \langle \phi_f + 2\pi k, T | \phi_i, 0 \rangle_{\mathbb{R}}
    =
    \sqrt{\frac{I}{2\pi i T}}
    \exp\!\left[
        \frac{iI(\phi_f - \phi_i + 2\pi k)^2}{2 T}
    \right].
\end{equation}
where we need to include the contribution of all winding sectors, $\phi(T) = \phi_f + 2\pi k$. Substituting this result into eq.~\eqref{eq:propagator_circ}, the propagator on the ring becomes
\begin{align}
    K(\phi_f,T;\phi_i,0)
    &=
    \sqrt{\frac{I}{2\pi i T}}
    \exp\!\left[\frac{i\theta}{2\pi}(\phi_f - \phi_i)\right]
    \sum_{k=-\infty}^{\infty}
    e^{ik\theta}
    \exp\!\left[
        \frac{iI(\phi_f - \phi_i + 2\pi k)^2}{2 T}
    \right] \nonumber\\
    &= \sqrt{\frac{I}{2\pi i T}} \exp\left( \frac{iI(\phi_f-\phi_i)^2}{2 T} + \frac{I\theta(\phi_f-\phi_i)}{2\pi} \right) \vartheta\left(\frac{I(\phi_f-\phi_i)}{ T} + \frac{\theta}{2\pi} ; \frac{2 I \pi}{ T} \right).
    \label{eq:propagator_PI}
\end{align}
By using the modular transformation, $\vartheta(z/\tau; -1/\tau) = (-i\tau)^{1/2} \exp\left( i\pi z^2/\tau \right) \vartheta(z;\tau)$, we can rewrite the propagator in eq.~\eqref{eq:propagator_Sch} as
\begin{align}
    K(\phi_f,T;\phi_i,0)|_{\rm eq.~\eqref{eq:propagator_Sch}}
    =& \sqrt{\frac{I}{2\pi i T}} \exp\left( \frac{iI(\phi_f-\phi_i)^2}{2 T} + \frac{I\theta(\phi_f-\phi_i)}{2\pi} \right) \nonumber\\
    & \times \vartheta\left(-\frac{I(\phi_f-\phi_i)}{ T} - \frac{\theta}{2\pi} ; \frac{2 I \pi}{ T} \right).
\end{align}
The equivalence between eqs.~\eqref{eq:propagator_Sch} and \eqref{eq:propagator_PI} can be verified using $\vartheta(z;\tau) = \vartheta(-z;\tau)$.

Here we have focused on the propagator $K(\phi_f,T;\phi_i,0)$ with finite $T$ and we can see that eq.~\eqref{eq:propagator_PI} cannot be equivalent to eq.~\eqref{eq:propagator_Sch} if we take a summation over $k$ with a finite upperbound in eq.~\eqref{eq:propagator_PI}.
Importantly, our results are consistent with the remark in Schulman’s textbook~\cite{Schulman:1981vu} that indicates for a particle on a ring, the propagator, $K(\phi_f,T;\phi_i,0)$, must be a periodic function of both the initial and final angles, $\phi_i$ and $\phi_f$. This periodicity, and hence the equivalence between the canonical and path-integral representations of the propagator, is guaranteed only when contributions from all integer winding sectors are included. Prescriptions that effectively restrict the sum over winding numbers therefore do not reproduce the full quantum-mechanical propagator.

\subsection{The partition function and energy spectra}\label{sec:quantum_rotor_energy}
The partition function of the quantum rotor is defined in Euclidean time as
\begin{equation}
Z(\theta, \beta) \equiv \mathrm{Tr}\!\left( e^{-\beta H } \right),
\label{eq:Z_def}
\end{equation}
where $\beta$ denotes the Euclidean time extent. This function can be evaluated equivalently using canonical quantization or the path-integral formalism.  
In the canonical approach, inserting a complete set of energy eigenstates of the Hamiltonian, one finds
\begin{equation}
Z(\theta, \beta) = \sum_n \langle n | e^{-\beta H } | n \rangle = \sum_n e^{-\beta E_n(\theta) }.
\label{eq:partition_function1}
\end{equation}
This representation makes the fact that the energy spectrum can be extracted from the exponential decomposition of $Z(\theta, \beta)$ explicit. Alternatively, the same quantity can be computed using the Euclidean path integral. For the quantum rotor, the Euclidean action in the presence of a $\theta$ term is
\begin{equation}
S_E[\phi;\theta] =
    \frac{I}{2} \int_0^{\beta} d\tau \, \dot{\phi}^2(\tau) - i\theta \int_0^{\beta} d\tau \, q(\tau),
    \label{eq:Euclidean_action}
\end{equation}
where $q(\tau)=\dot{\phi}(\tau)/2\pi$ is the topological charge density as it is defined in ref.~\cite{Ai:2025quf}.

\medskip
Then, the partition function takes the form 
\begin{align}
Z(\theta, \beta) = \int \mathcal{D}\phi \, \exp \Big( - S_E[\phi;\theta] \Big) &= \sqrt{\frac{2\pi I}{  \beta}}  \sum_{n\in\mathbb{Z}}  \exp\left[- \frac{2 \pi^2 I }{ \beta} \, n^2 + i n \theta \right] \notag\\
&= \sum_{n\in\mathbb{Z}} \exp\!\left[ -\frac{ \beta}{2I} \left(n+\frac{\theta}{2\pi}\right)^2 \right],
\label{eq:partition_function2}
\end{align}
where we have used the results from section~\ref{sec:propagatorsec} in the first line and we have used
the modular transformation $\vartheta(z/\tau; -1/\tau) = (-i\tau)^{1/2} \exp\left( i\pi z^2/\tau \right) \vartheta(z;\tau)$ with $\tau = 2\pi i I  \beta$ and $z = \theta/2\pi$
in the second line.
This way, we can explicitly demonstrate that the path integral formalism can reproduce the results of the canonical quantization when the contribution of all winding sectors is summed over. 

We can use this framework to examine the order of limits discussed by ACGT. Consider instead a truncated partition function in which only a finite range of winding numbers is included. This truncation should be viewed purely as a technical device to make the order of limits explicit, and its rationale and interpretation are discussed in detail in section~\ref{sec:Top_Susceptibility}. So, let's define a truncated partition function as,
\begin{equation}
Z_{\Delta}(\theta) = \sqrt{\frac{2\pi I}{ \beta}} \sum_{n=-\Delta}^{\Delta} \exp\!\left[in\theta - \frac{2\pi^2 I n^2}{ \beta} \right],
\label{eq:Z-truncated}
\end{equation}
where $\Delta\in\mathbb{N}$ is held fixed. For fixed $\Delta$ and large $\beta$, the Gaussian factor varies slowly over $|n|\leq\Delta$ and may be neglected at leading order,
\begin{equation}
Z_{\Delta}(\theta) \;\xrightarrow[\beta\to\infty]{}\; \sqrt{\frac{2\pi I}{ \beta}} \sum_{n=-\Delta}^{\Delta} e^{in\theta} = \sqrt{\frac{2\pi I}{ \beta}} \frac{\sin\!\left[(\Delta+\tfrac12)\theta\right]}{\sin(\theta/2)} .
\label{eq:Z-truncated-largebeta}
\end{equation}
This truncated expression fails to reproduce the correct energy spectrum of the quantum rotor. Moreover, the prefactor $\beta^{-1/2}$, is analogous to the $(VT)^{-1/2}$ factor in ref.~\cite{Khoze:2025auv}.

From eq.~\eqref{eq:partition_function1}, for $\beta \to \infty$ the partition function is dominated by the ground state, and one defines
\begin{equation}
E_0 = -\lim_{\beta\to\infty} \frac{1}{\beta}\log Z(\theta, \beta).
\label{eq:gs_energy}
\end{equation}
Using the truncated partition function, however, one obtains
\begin{equation}
E_0 \sim \frac{1}{\beta}\log Z_{\Delta} \;\xrightarrow[\beta\to\infty]{}\; 0 \qquad (\text{for fixed } \Delta). \label{eq:E0 from fixed topology}
\end{equation}
This result is equivalent to the claim by ACGT \cite{Ai:2020ptm, Ai:2025quf} that $\theta$-dependence in the partition function enters only through a normalization factor, $Z_\theta = N(\theta) Z_0$. However, it is in clear disagreement with the well‑known and solid results in quantum mechanics.

\subsection{Topological susceptibility}
\label{sec:Top_Susceptibility}
We now discuss the topological susceptibility of the quantum rotor and reexamine the discussions by Albandea \emph{et al.}~\cite{Albandea:2024fui}, Benabou \emph{et al.}~\cite{Benabou:2025viy}, and Ai \emph{et al.}~\cite{Ai:2025quf}. 

In the standard treatment of Yang--Mills theory, and in particular in the Witten--Veneziano framework, the topological susceptibility is fundamentally defined as the curvature of the vacuum energy density with respect to the $\theta$ angle,
\begin{equation}
    \chi \equiv \frac{\partial^2}{\partial \theta^2} E_0(\theta) \bigg|_{\theta=0}
    = -\lim_{\beta\to\infty} \frac{1}{\beta}\left. \frac{\partial^2}{\partial \theta^2} \Big( \log Z(\theta, \beta) \Big) \right|_{\theta=0}.
    \label{eq:chi_def}
\end{equation}
Using $Q=\int d\tau\, q(\tau)$, one may rewrite this as
\begin{equation}
    \chi_{\rm glob} = \lim_{\beta\to\infty} \frac{1}{\beta} \int d\tau\, d\tau'\, \langle q(\tau) q(\tau') \rangle = \lim_{\beta\to\infty}\frac{1}{\beta}\langle Q^2\rangle.
    \label{eq:chi_global}
\end{equation}
If the vacuum is translationally invariant, and taking the infinite-volume limit yields the equivalent representation
\begin{equation}
    \chi_{\rm loc} = \lim_{\beta\to\infty} \int d\tau\, \langle q(\tau) q(0)\rangle .
    \label{eq:chi_local}
\end{equation}
Equations~\eqref{eq:chi_global} and~\eqref{eq:chi_local} are referred to as the ``global'' and ``local'' representations of the susceptibility, respectively. In the standard formulation of the theory, these two expressions should coincide as a consequence of translational invariance. Thus, they are simply different representations of the same quantity. Indeed, we can see this equivalence when the path integral includes the sum over all topological sectors before the infinite-volume limit is taken. Nevertheless, as thoroughly explained by Martin L\"uscher~\cite{Luscher:2004fu,Luscher:2010ik}, one needs to be very careful when using the so-called local definition as it contains a divergent contact term that might be very scheme dependent. This expectation is supported by lattice studies in which the topological susceptibility is defined nonperturbatively and shown to have a universal continuum limit once the short-distance contact term is treated properly.

The purpose of the discussion below is therefore not merely to compare
different formulas, but to test different prescriptions against the exact
benchmark.

\subsubsection*{Method A: Euclidean correlation function with semiclassical expansion}
Taking two derivatives of \eqref{eq:partition_function2} and evaluating the result at $\theta=0$, we obtain
\begin{equation}
    \chi = \lim_{\beta\to\infty} \frac{- 1}{\beta} \frac{1}{Z(0,\beta)} \int \mathcal{D}\phi \, \left( i \int_0^\beta d\tau \, q(\tau) \right) \left( i \int_0^\beta d\tau' \, q(\tau') \right) e^{-S_E[\phi;0] } .
    \label{eq:chi_path_integral}
\end{equation}
This equation can be rewritten as an expectation value of the squared integrated topological charge, $\chi = \beta^{-1}\langle Q^2\rangle$ with $Q=\int_0^\beta d\tau\, q(\tau)$ as it is done in ref.~\cite{Benabou:2025viy}. Moreover, using time-translation invariance, and taking the integral over $\tau'$, it can be written as $\chi = \int_0^\beta d\tau\,\langle q(\tau) q(0)\rangle$, as in eq.~(1) of ref.~\cite{Ai:2025quf}. However, in this form the $\beta\to\infty$ limit is implicit, and its interplay with the temporal integrations is not manifest. As we show below, the two prescriptions are equivalent only when the order of limits is specified. 

In contrast to the approach adopted in ref.~\cite{Ai:2025quf}, here, we will proceed by evaluating the path integral directly. This is quite straight forwards as $\int d\tau \, q(\tau) = \int d\tau \; \dot{\phi}(\tau)/2\pi = n\in\mathbb{Z}$. Therefore the susceptibility is simply,
\begin{align}
    \chi = \lim_{\beta\to\infty}\frac{1}{\beta} \frac{1}{Z(0,\beta)} \int \mathcal{D}\phi \, n^2 \, e^{-S_E[\phi;0] } = \lim_{\beta\to\infty} \frac{1}{\beta} \frac{ \sum_{n\in\mathbb{Z}} n^2 \, \exp\!\big( -\frac{2\pi^2 I}{ \beta} n^2 \big) }{ \sum_{n\in\mathbb{Z}} \exp\!\big( -\frac{2\pi^2 I}{ \beta} n^2 \big) }.
\end{align}
which is equivalent to the eq.~(37) in ref.~\cite{Benabou:2025viy}. To connect with the correlator-based approach of ref.~\cite{Ai:2025quf}, we return to eq.~\eqref{eq:chi_path_integral} and write
\begin{equation}
    \chi
    =
    \lim_{\beta\to\infty}
    \frac{1}{4\pi^2\beta}
    \int_0^\beta d\tau\, \int_0^\beta d\tau'\,
    \frac{1}{Z(0,\beta)}
    \int\!\mathcal{D}\phi\,
    \dot{\phi}(\tau)\dot{\phi}(\tau')\,
    e^{-S_E[\phi;0]} .
\end{equation}
We now expand the path integral semiclassically around classical winding solutions,
\begin{equation}
    \phi(\tau)=\phi_{\rm cl}(\tau)+\eta(\tau),
    \qquad
    \phi_{\rm cl}(\tau)=\phi_0+\frac{2\pi n}{\beta}\tau,
    \qquad
    \eta(0)=\eta(\beta)=0 .
    \label{eq:phi_decomposition}
\end{equation}
As a result, the measure factorizes into winding and fluctuation parts and substituting $\dot{\phi}(\tau)$,
\begin{align}
    \chi &= \lim_{\beta\to\infty} \frac{1}{4\pi^2 \beta} \int_0^\beta d\tau \int_0^\beta d\tau' \, \sum_{n\in\mathbb{Z}} \frac{1}{Z(0,\beta)} \int \mathcal{D}\eta \, \left( \frac{2\pi n}{\beta} + \dot{\eta}(\tau) \right)
    \left( \frac{2\pi n}{\beta} + \dot{\eta}(\tau') \right) \notag \\
    &\quad \times \exp\!\left[ -\frac{I}{2} \int_0^\beta d\tau'' \, \dot{\eta}^2 \right]
    \exp\!\left[ -\frac{2\pi^2 I}{ \beta} n^2 \right].
\end{align}
The cross terms between the classical configuration and the fluctuation field vanish identically due to the Dirichlet boundary conditions imposed on $\eta$. As a result, the topological susceptibility can be written as
\begin{align}
    \chi &= \lim_{\beta\to\infty}\frac{1}{4\pi^2 \beta} \Bigg[\frac{ \int d\tau  d\tau'\, \sum_{n\in\mathbb{Z}} \, \left( \frac{2\pi n}{\beta} \right)^2 \, \exp\!\left[ -\frac{2\pi^2 I}{ \beta} n^2 \right] }{ \sum_{n\in\mathbb{Z}} \exp\!\left[ -\frac{2\pi^2 I}{ \beta} n^2 \right] } \nonumber\\
    &\qquad\qquad\qquad~
    + \frac{\int d\tau  d\tau'\, \int \mathcal{D}\eta \; \dot{\eta}(\tau) \dot{\eta}(\tau')\, \exp\!\left[ -\frac{I}{2} \int d\tau'' \, \dot{\eta}^2 \right] }{ \int \mathcal{D}\eta \, \exp\!\left[ -\frac{I}{2} \int d\tau'' \, \dot{\eta}^2 \right] }\Bigg]
\end{align}
Using the identity,
\begin{align}
    &\frac{1}{Z[0]}\int {\cal D}\eta~\dot \eta(\tau) \dot \eta(\tau') e^{-S_E[\eta;0]} = \frac{d}{d\tau} \frac{d}{d\tau'} \frac{1}{Z[0]}\int {\cal D}\eta~ \eta(\tau) \eta(\tau') e^{-S_E[\eta;0]} \equiv \frac{d}{d\tau} \frac{d}{d\tau'} \langle {\rm T }\, \eta(\tau) \eta(\tau') \rangle.
\end{align}
The susceptibility simplifies to
\begin{align}
    \chi = \lim_{\beta\to\infty} \left[ \frac{\langle n^2\rangle}{\beta} + \frac{1}{4\pi^2\beta} \int_0^\beta d\tau\, \int_0^\beta d\tau'\, \frac{d}{d\tau}\frac{d}{d\tau'} \langle {\rm T}\, \eta(\tau) \eta(\tau') \rangle \right].
    \label{eq:chi_4}
\end{align}
Then, defining the fluctuation propagator as $G(\tau,\tau'; \beta)=\langle\eta(\tau)\eta(\tau')\rangle$ that 
satisfies Dirichlet boundary conditions, $G(0,\tau'; \beta)=G(\beta,\tau'; \beta)=0,$ we obtain (see, e.g., chapter 3 of \cite{Kleinert:2004ev})
\begin{align}
    G(\tau,\tau'; \beta) = \frac{\tau + \tau' - |\tau-\tau'|}{2I} - \frac{\tau \tau'}{\beta I},
\end{align}
So,
\begin{equation}
    \frac{d}{d\tau}\frac{d}{d\tau'} G(\tau,\tau'; \beta) = \frac{1}{I}\,\delta(\tau-\tau') - \frac{1}{\beta I}.
\end{equation}
Substituting this result in eq. \eqref{eq:chi_4}, we obtain
\begin{align}
    \chi = \lim_{\beta\to\infty} \left[ \frac{\langle n^2\rangle}{\beta} + \frac{1}{4\pi^2\beta} \int_0^\beta d\tau\, \int_0^\beta d\tau' \left( \frac{1}{I}\,\delta(\tau-\tau') - \frac{1}{\beta I} \right) \right].
    \label{eq:chi_5}
\end{align}
Eq.~\eqref{eq:chi_5} makes explicit that the Gaussian fluctuation contribution vanishes identically for any finite $\beta$, due to an exact cancellation between the local $\delta$-function term and the $\beta$-dependent boundary contribution imposed by Dirichlet conditions.
The nonzero result reported in ref.~\cite{Ai:2025quf} for a fixed topological sector stems from interchanging the limit $\beta\to\infty$ with the temporal integrations which is not valid as the integrand contains a $\beta$-dependent constant whose integral exactly cancels the $\delta$-function contribution.

Moreover, eq.~\eqref{eq:chi_5} makes the validity of our argument explicit, as the susceptibility is evaluated without assuming time-translation invariance. In this formulation, the Gaussian contribution vanishes regardless of the order in which infinite-time limit and temporal integrations are performed. A nonzero susceptibility emerges only upon summing over all winding sectors, demonstrating that the result is a genuine consequence of the full topological sum rather than an artifact of integration order or symmetry assumptions.

\subsubsection*{Method B: Euclidean correlation function of momentum operator}
The derivative of the partition function $Z$ given in eq.~\eqref{eq:chi_def} can be written as the correlation function of the momentum operators.
Promoting $\theta$ to a $\tau$-dependent source $\theta(\tau)$, functional derivatives of $\log Z$ generate correlation functions of the momentum operator. 
Then, eq.~\eqref{eq:chi_path_integral} can be rewritten as
\begin{align}
    \chi = \lim_{\beta\to\infty} \frac{- 1}{\beta} \frac{1}{Z(0,\beta)}
    \int_0^\beta d\tau \int_0^\beta d\tau' 
    \frac{\delta}{\delta \theta(\tau)}\frac{\delta}{\delta \theta(\tau')} Z \biggr|_{\theta(\tau)=0}
\end{align}
and we find,
\begin{align}
    \frac{\delta}{\delta \theta(\tau)}\frac{\delta}{\delta \theta(0)} \log Z \biggr|_{\theta(\tau)=0}
    &=
    \left[ -\left( \frac{1}{Z} \frac{\delta Z}{\delta \theta(0)} \right)^2
    + \frac{1}{Z} \frac{\delta}{\delta \theta(\tau)}\frac{\delta}{\delta \theta(0)} Z \right]_{\theta(\tau)=0} \nonumber\\
    &= -\frac{1}{4\pi^2 I^2} \langle 0|p |0\rangle^2 - \frac{1}{4\pi^2 I}\delta(\tau) + \frac{1}{4\pi^2 I^2} \langle 0|T p(\tau) p(0) |0\rangle, \label{eq:diff of chi}
\end{align}
where $p(\tau) = e^{H\tau} p(0) e^{-H\tau}$ and $|0\rangle$ denotes the ground state. Using the same prescription for extracting $\chi$ as in the previous subsection, we integrate over $\tau$ to obtain
\begin{align}
    \chi \big|_{\theta=0}
    &=
    \frac{1}{4\pi^2 I}
    -
    \frac{1}{4\pi^2 I^2}
    \int_{-\infty}^{\infty} d\tau
    \left[
        \langle 0 | T p(\tau) p(0) | 0 \rangle
        - \langle 0 | p | 0 \rangle^2
    \right].
    \label{eq:ts from pi}
\end{align}
Using the spectral representation of the Euclidean correlator,
\begin{align}
    \int_{-\infty}^\infty d\tau \left[ \langle 0| Tp(\tau) p(0)|0\rangle - \langle 0| p|0 \rangle^2 \right]
    &= 2 \int_0^\infty d\tau \left[ \sum_n \langle 0|p(\tau) |n\rangle \langle n|  p(0)|0\rangle - \langle 0|p |0\rangle^2 \right] \nonumber\\
    &= 2 \int_0^\infty d\tau \sum_{n\neq 0} \left[ |\langle n| p(0)|0\rangle|^2 \exp\left( -(E_n-E_0)\tau\right) \right] \nonumber\\
    &= 2 \sum_{n\neq 0} \frac{1}{E_n-E_0} |\langle n| p(0)|0\rangle|^2. \label{eq:ts from pi 2}
\end{align}
we recover
\begin{align}
    \chi \big|_{\theta=0} = \frac{1}{4 \pi^2 I} - \frac{1}{2 \pi^2 I^2} \sum_{n\neq 0} \frac{|\langle n | p | 0 \rangle|^2}{E_n(0) - E_0(0)}. \label{eq:ts from methodB}
\end{align}
Since the ground state of the quantum rotor satisfies $p|0\rangle = 0$, we find that this result in exact agreement with Method A.

\subsubsection*{Method C: Ground-state energy}
In the above two method, we have calculated $\chi$ from the Euclidean path integral formulation. We can also derive the equivalent results from $\theta$-dependence of the spectrum of the system.
For fixed $\theta$, the system has a discrete spectrum $E_n(\theta)$.
In large $\beta$ limit, $\log Z(\theta,\beta)$ behaves as $-\beta E_0(\theta)$.
Using the definition of the susceptibility already introduced in
eq.~\eqref{eq:chi_def}, we may write
\begin{align}
    \chi \big|_{\theta=0}
    =
    \left.
    \frac{\partial^2 E_0(\theta)}{\partial \theta^2}
    \right|_{\theta=0}.
    \label{eq:ts from e0}
\end{align}
For small $\theta$, $E_0(\theta)$ can be calculated by using perturbative expansion as
\begin{align}
    E_0(\theta)
    =
    E_0(0)
    + \frac{\theta}{2I\pi} \langle 0 |p|0 \rangle
    +  \theta^2 \left[ \frac{1}{8I\pi^2} - \frac{1}{4I^2\pi^2} \sum_{n\neq 0} \frac{|\langle n|p|0\rangle|^2}{E_n(0) - E_0(0)} \right] + \cdots,
\end{align}
Taking two derivatives with respect to $\theta$ and evaluating the result at $\theta=0$, we obtain
\begin{align}
    \chi|_{\theta=0} = \frac{1}{4\pi^2 I} - \frac{1}{2\pi^2 I^2} \sum_{n\neq 0} \frac{|\langle n|p|0\rangle|^2}{E_n(0) - E_0(0)}. \label{eq:ts from e0 2}
\end{align}
We obtained the result equivalent to eq.~\eqref{eq:ts from methodB}.
Again, by using $p|0\rangle = 0$, we obtain $\chi = 1/4\pi^2 I$.

\subsubsection*{Comments on the ACGT Order-of-Limits}

We have demonstrated that all three approaches yield the same result for the topological susceptibility, $\chi = 1/4\pi^2 I$. This result, together with its interpretation, is in full agreement with the findings of Albandea \emph{et al.}~\cite{Albandea:2024fui} and Benabou \emph{et al.}~\cite{Benabou:2025viy}. Now let us discuss the implications on ACGT proposal on the order of the limits in the path integral formulation and the discussion on the topological susceptibility in ref.~\cite{Ai:2025quf}.

First, we introduce a technical device that allows us to address the order-of-limits issue in a direct and transparent way and to pinpoint the origin of the nonvanishing susceptibility in the full path-integral formulation. Specifically, we impose a finite cutoff $\Delta$ on the winding number. This cutoff allows us to cleanly distinguish between the order in which the summation over all winding sectors $(\Delta \to \infty)$ and the infinite-time limit $(\beta \to \infty)$ are taken. The same strategy was employed in the analysis of the propagator (section~\ref{sec:propagatorsec}) and of the partition function and energy spectrum (section~\ref{sec:quantum_rotor_energy}), and will be used again in the next section for the quantum pendulum. We emphasize that truncating the winding sum is purely a technical device, introduced solely to test and clarify the order-of-limits prescription proposed by ACGT.

We now return to eq.~\eqref{eq:chi_5}, where the nonzero contribution to $\chi$ appears to arise from winding configurations. Introducing the finite cutoff $\Delta$, eq.~\eqref{eq:chi_5} can be rewritten as
\begin{align}
    \chi(\Delta) = \lim_{\Delta\to\infty}\lim_{\beta\to\infty}\frac{1}{\beta}\, \frac{\sum_{|n|\leq \Delta} n^2  \exp\!\left(-\frac{2\pi^2 I}{\beta}n^2\right)} {\sum_{|n|\leq \Delta}  \exp\!\left(-\frac{2\pi^2 I}{\beta}n^2\right)}, 
    \label{eq:chiDelta}
\end{align}
Here \(\Delta\) denotes the maximal winding number allowed to contribute to the path integral. For any fixed, finite $\Delta$, the explicit $1/\beta$ prefactor suppresses the result as \(\beta\to\infty\), and $\chi(\Delta)=0$. In contrast, performing the winding sum first yields
\begin{equation}
    \chi = \lim_{\beta\to\infty} \lim_{\Delta\to\infty} \chi(\Delta) = \frac{1}{4\pi^2 I},
    \label{eq:chiFinal}
\end{equation}
This result is in full agreement with standard quantum-mechanical expectations, lattice simulations, and ref.~\cite{Benabou:2025viy}.
In particular, as shown in method C, the ground-state energy of the quantum rotor in the presence of a $\theta$ term is $E_0(\theta)=\frac{1}{2I}\left(\frac{\theta}{2\pi}\right)^2$, which in turn implies a nonvanishing topological susceptibility, $\chi=\left.\partial_\theta^2 E_0\right|_{\theta=0}=1/(4\pi^2 I)$.
The conclusion is therefore unambiguous: if the limit $\beta\to\infty$ is taken at fixed winding number (or within any finite subset of winding sectors), the susceptibility vanishes, whereas a nonzero topological susceptibility arises only when the sum over all winding sectors is performed prior to the infinite-time limit. This feature lies at the core of the Euclidean path-integral formulation and reflects the necessity of including all allowed paths in the functional integral.

Second, our method~B is equivalent to the calculation of $\chi$ in eq.~(4) of Ai \emph{et al.}~\cite{Ai:2025quf}.
For the quantum rotor, $\langle 0 |p|0\rangle = 0$ and $\langle 0 | T p(\tau) p(0)|0\rangle = 0$ are satisfied because of $p|0\rangle = 0$. Thus, eq.~\eqref{eq:diff of chi} tells us that
\begin{align}
    \frac{\delta}{\delta \theta(\tau)}\frac{\delta}{\delta \theta(0)} \log Z
    &= \frac{1}{4\pi^2 I^2}\delta(\tau),
\end{align}
and we obtain
\begin{align}
    \chi = \int_{-\infty}^\infty d\tau \frac{\delta}{\delta \theta(\tau)}\frac{\delta}{\delta \theta(0)} \log Z
    = \frac{1}{4\pi^2 I}.
\end{align}
Although we have explicitly verified the correctness of the calculation of $\chi$ presented in ref.~\cite{Ai:2025quf}, this agreement at the level of the final result does not, by itself, justify the ACGT order-of-limits prescription. The reason is that the order of the limits in the path integral does not appear directly in ref.~\cite{Ai:2025quf} calculation.
The crucial point is that both our analysis in Method B and that of ref.~\cite{Ai:2025quf} rely on an implicit assumption about the ground state, namely $p|0\rangle = 0$, which is sufficient to obtain $\chi = 1/(4\pi^2 I)$, independently of any discussion of the path integral. In particular the condition $p|0\rangle = 0$ immediately implies $\exp\left( 2 \pi i p \right) |0\rangle = |0\rangle$. Since $p$ is the generator of translations, the operator $e^{2\pi i p}$ implements a $2\pi$ shift of the angular variable, corresponding to a large gauge transformation in the quantum rotor. Thus, assuming $p|0\rangle = 0$ is equivalent to selecting a large-gauge-invariant ground state. If such a gauge-invariant state is imposed as a boundary condition in the path-integral formulation, the functional integral necessarily includes a sum over all topological sectors. Consequently, the calculation of the topological susceptibility in ref.~\cite{Ai:2025quf}, while correct, is not equivalent to the path-integral prescription proposed by ACGT refs.~\cite{Ai:2020ptm, Ai:2024cnp, Ai:2025quf}. The latter requires a gauge-variant boundary condition at finite spacetime volume, with gauge invariance recovered only after taking the $VT \to \infty$ limit.

Third, we comment on the order of limit $\beta \to \infty$ and the temporal integration. Ref.~\cite{Ai:2025quf} claimed that $\chi = 1/(4\pi^2 I)$ can be obtained even within a fixed topological sector; however, this derivation relies on interchanging the $\beta\to\infty$ limit with the temporal integration as
\begin{align}
    \chi = \int_{-\infty}^\infty d\tau \left( \lim_{\beta\to\infty} \frac{1}{4\pi^2}\frac{d}{d\tau} \frac{d}{d\tau'} G(\tau,\tau';\beta)   \right),
\end{align}
which is not justified as the integrand contains a $\beta$-dependent constant whose integral exactly cancels the $\delta$-function contribution.
As shown explicitly in eq.~\eqref{eq:chi_5} in the method A, the fluctuation contribution to $\chi$ vanishes identically for any finite $\beta$, leaving only the winding-sector term $\langle n^2\rangle/\beta$, which goes to zero as $\beta\to\infty$ for any fixed winding number.

The rotor provides an instructive example in which the local and global representations of the susceptibility agree in the physical vacuum, but for different-looking reasons. In the global formulation, the nonzero result arises from the unrestricted sum over winding sectors, which reproduces the exact $\theta$-dependence of the ground-state energy. In the local formulation, the same value arises as a contact term fixed by the canonical commutation relations, as in Method~B. These are not competing definitions of different observables; rather, they are two representations of the same susceptibility in the correctly defined vacuum. The apparent tension arises only if one takes the infinite-time limit before summing over all topological sectors. In that case the global expression vanishes, but this vanishing susceptibility no longer reproduces the exact $\theta$-dependence of the energy spectrum and therefore does not correspond to the physical susceptibility.

Finally, we emphasize that nonzero $\chi$ directly means that $\theta$ is a physical parameter because it affects $E_0(\theta)$, which is a physical observable. As illustrated in the appendix \ref{sec:charged particle}, nonzero $\theta$ on a ring can be understood as the Aharonov-Bohm effect due to a magnetic flux inside the ring. Also, for small $\theta$, we can show that $\langle 0|p|0 \rangle \simeq \theta \chi$, which is analogous to $\langle (g_s^2/32\pi^2) G_{\mu\nu} \tilde G^{\mu\nu} \rangle \simeq \theta \chi_t$ in QCD \cite{Shifman:1979if}.


\section{A ring with potential: quantum pendulum}\label{sec:quantum-pendulum}
In the previous section, we have discussed quantum rotor system which is equivalent to a particle on a ring without potential.
In this section, we discuss the system with nonzero potential. This potential can be either electric potential or gravitational potential. A particle on a ring with gravitational potential can be interpreted as a quantum pendulum, which has been discussed in the literature such as refs.~\cite{Rubakov:2002fi, Bachas:2016ffl}.

Now the Schr\"odinger equation of this system is given as
\begin{align}
	-\frac{1}{2I} \frac{\partial^2\psi}{\partial\phi^2} + V(\phi) \psi(\phi) = E\psi(\phi) \label{eq:schroedinger eq with V}
\end{align}
with the twisted boundary condition for $\psi$:
\begin{align}
	\psi(\phi + 2\pi) = e^{i\theta} \psi(\phi). \label{eq:twist bdry condtion}
\end{align}
Without loss of generality, we can assume $V(\phi)$ has the global minimum at $\phi=0$ and $V(\phi)$ around the minimum can be expanded as
\begin{align}
	V(\phi) \simeq \frac{1}{2} I \omega^2 \phi^2.
\end{align}
The ground state wave function is localized around $\phi\simeq 2\pi n$ with integer $n$, and the wave function should satisfy the boundary condition eq.~\eqref{eq:twist bdry condtion} at the same time. Thus, the construction of the ground state in this system is quite similar to $\theta$-vacua in gauge theories. See, e.g., \cite{Rubakov:2002fi}.

\subsection{The partition function}
As section \ref{sec:quantum-rotor}, we calculate the partition function $Z$.
The partition function $Z$ is defined as similar to eq.~\eqref{eq:partition_function1}, and we obtain
\begin{align}
    Z &= \sum_n e^{-E_n \beta }.
\end{align}
Let us calculate $Z$ by using two different methods.

\subsubsection*{Method A: Schr\"odinger equation with WKB approximation}
First, let us solve Schr\"odinger equation \eqref{eq:schroedinger eq with V} directly.
Since the ground state gives the dominant contribution for large $\beta$ as $Z \simeq \exp(-E_0 \beta),$ we calculate the energy of the ground state $E_0$ by utilizing the WKB approximation. To this end, it is useful to define the following dimensionless variables:
\begin{align}
	z \equiv \sqrt{I\omega} \phi, \qquad
	v(z) \equiv \frac{V( \sqrt{  1/I\omega}z)}{\omega}, \qquad
	\nu \equiv \frac{E}{\omega} - 1/2.
\end{align}
$v(z)$ is a periodic function as $v(z) = v(z+p)$ where the period $p$ is
\begin{align}
	p = 2\pi\sqrt{I\omega}.
\end{align}
Then, we obtain the reduced Schr\"odinger equation and boundary condition as
\begin{align}
	-\frac{1}{2} \psi''(z) + v(z) \psi(z) = \left( \frac{1}{2} + \nu \right) \psi(z), \qquad
    \psi(z + p) = e^{i\theta} \psi(z).
\end{align}
The reduced potential $v(z)$ has minima at $z = np$ with integer $n$, corresponding to a periodic array of identical wells. We assume that the potential is sufficiently deep and smooth so that, in the vicinity of each minimum, it may be approximated by a harmonic oscillator. More precisely, there exists a small parameter $\epsilon$ such that
\begin{equation}
    v(z) \simeq \frac{(z-np)^2}{2},
    \qquad np-\epsilon < z < np+\epsilon ,
\end{equation}
while in the region $np+\epsilon \le z \le (n+1)p-\epsilon$ the potential is large enough to justify the WKB approximation. In this regime, the low-energy states are localized near individual minima, and for the ground state the parameter $\nu$ appearing in the parabolic cylinder functions is expected to satisfy $\nu \simeq 0$. Without loss of generality, we focus on the minimum at $z=0$.

Hence, in the region $-\epsilon < z < \epsilon$, the wavefunction can be written as a linear combination of parabolic cylinder functions,
\begin{equation}
	\psi(z) \equiv \xi\, D_\nu(-\sqrt{2}\,z) + (1-\xi)\, D_\nu(\sqrt{2}\,z),
	\label{eq:psi at z=0}
\end{equation}
where $\xi$ is a coefficient to be fixed by the boundary condition. For $\nu \simeq 0$, the asymptotic behavior of \eqref{eq:psi at z=0} is
\begin{align}
	\psi(z) &\to e^{-z^2/2} + (1-\xi) \frac{\nu \sqrt{\pi}}{z} e^{z^2/2}, & (z\to-\infty) \label{eq:psi -infty} \\
	\psi(z) &\to e^{-z^2/2} - \xi \frac{\nu \sqrt{\pi}}{z} e^{z^2/2}. & (z\to+\infty) \label{eq:psi +infty}
\end{align}
Using \eqref{eq:psi -infty}, the wavefunction in the region $-p+\epsilon \le z \le -\epsilon$ can be matched to $\psi(-\epsilon)$ via the WKB approximation,
\begin{align}
	\psi(z) \simeq  \left( \frac{v(z) - 1/2}{v(-\epsilon) - 1/2} \right)^{1/4} \Biggl[ 
        & e^{-\epsilon^2/2} \exp\left( \int_{-\epsilon}^z dz' \sqrt{ 2 v(z') - 1} \right) \nonumber\\
		& + (1-\xi) \frac{\nu\sqrt{\pi}}{-\epsilon} e^{\epsilon^2/2} \exp\left( - \int_{-\epsilon}^z dz' \sqrt{ 2 v(z') - 1} \right)
	\Biggr]. \label{eq:psi negative z}
\end{align}
Similarly, using \eqref{eq:psi +infty}, the wavefunction in the region $\epsilon \le z \le p-\epsilon$ is
\begin{align}
	\psi(z) \simeq \left( \frac{v(z) - 1/2}{v(\epsilon) - 1/2} \right)^{1/4} \Biggl[ 
        & e^{-\epsilon^2/2} \exp\left( -\int_\epsilon^z dz' \sqrt{ 2 v(z') - 1} \right) \nonumber\\
		& -\xi \frac{\nu\sqrt{\pi}}{\epsilon} e^{\epsilon^2/2} \exp\left( + \int_\epsilon^z dz' \sqrt{ 2 v(z') - 1} \right)
	\Biggr]. \label{eq:psi positive z}
\end{align}
To determine the tunneling-induced energy splitting, we impose the twisted boundary condition, $\psi(z + p) = e^{i\theta} \psi(z)$. For this purpose, it is convenient to rewrite \eqref{eq:psi positive z} as
\begin{align}
	\psi(z) \simeq \left( \frac{v(z) - 1/2}{v(\epsilon) - 1/2} \right)^{1/4} \Biggl[ 
        & k e^{-S} \frac{\sqrt{\pi}}{\epsilon} e^{\epsilon^2/2} \exp\left( -\int_{-\epsilon+p}^z dz' \sqrt{ 2 v(z') - 1} \right) \nonumber\\
		& -\frac{\xi \nu}{k e^{-S}} e^{-\epsilon^2/2} \exp\left( + \int_{-\epsilon+p}^z dz' \sqrt{ 2 v(z') - 1} \right)
	\Biggr], \label{eq:psi positive z2}
\end{align}
where $S$ and $k$ are defined as
\begin{align}
	S &\equiv \int_0^p dz' \sqrt{2v(z')}, \label{eq:S from Schroedinger eq}\\
	k &\equiv \frac{\epsilon}{\sqrt{\pi}} \exp\left( -\epsilon^2 + \int_0^p dz' \sqrt{2v(z')} - \int_\epsilon^{-\epsilon + p} dz' \sqrt{2v(z')-1}\right). \label{eq:K from Schroedinger eq}
\end{align}
Solving the twisted boundary condition using \eqref{eq:psi negative z} and \eqref{eq:psi positive z2}, we obtain
\begin{equation}
	\nu = -2k e^{-S} \cos\theta,
	\qquad
	\xi = \frac{e^{i\theta}}{2\cos\theta}.
\end{equation}
Defining $K \equiv k\omega$, the ground-state energy takes the form
\begin{equation}
	E_0 = \frac{1}{2}\omega - 2K e^{-S} \cos\theta.
	\label{eq:E0 from schroedinger eq}
\end{equation}
This expression explicitly exhibits the tunneling-induced band structure. As we will show in the next subsection, the quantities $S$ and $K$ defined in
eqs.~\eqref{eq:S from Schroedinger eq} and \eqref{eq:K from Schroedinger eq}
coincide with the bounce action and the associated functional determinant
computed in the semiclassical path-integral formalism.

Using the ground-state energy defined as eq.~\eqref{eq:E0 from schroedinger eq}, the partition function $Z$ for large $T$ behaves as
\begin{align}
    Z \simeq \exp\left( -\frac{1}{2}\omega \beta + 2 K \beta e^{-S} \cos\theta \right). \label{eq:Z from schroedinger eq}
\end{align}
where the $\theta$-dependence originates from tunneling between adjacent minima. Moreover, the topological susceptibility $\chi$ is calculated as 
\begin{align}
    \chi
    = -\lim_{\beta\to\infty} \frac{1}{\beta} \frac{\partial^2}{\partial\theta^2} \log Z
    = \frac{\partial^2 E_0}{\partial \theta^2}
    = 2 K e^{-S} \cos\theta. \label{eq:chi from schroedinger eq}
\end{align}
which shows that the susceptibility is entirely controlled by the tunneling amplitude and is exponentially suppressed in the semiclassical limit.

\subsubsection*{Method B: Path integral with dilute instanton gas approximation}

We now calculate the partition function $Z(\theta, \beta)$ from the Euclidean path integral, which provides a complementary perspective to the Schrödinger approach. For a finite Euclidean time interval $\beta$, the action is
\begin{align}
    S_E[\phi] = \int_0^\beta d\tau \, \Bigg[ 
        \frac{1}{2} I \left( \frac{d\phi}{d\tau} \right)^2 + V(\phi(\tau)) 
        + \frac{i\theta}{2\pi} \frac{d\phi}{d\tau} 
    \Bigg].
\end{align}
For large $\beta$, the path integral is dominated by configurations around classical solutions, i.e., saddle points of $S_E$. One such solution is the \emph{bounce} (instanton) $\phi_B(\tau)$, which satisfies the classical equation of motion and interpolates between physically equivalent vacua:
\begin{align}
    I \frac{d^2 \phi_B}{d\tau^2} = V'(\phi_B(\tau)), \qquad
    \lim_{\tau \to -\infty} \phi_B(\tau) = 0, \quad
    \lim_{\tau \to +\infty} \phi_B(\tau) = 2\pi.
\end{align}
with $\phi=0$ and $\phi=2\pi$ identified. Using the dilute instanton gas approximation~\cite{Coleman:1985rnk}, we can sum over configurations with arbitrary numbers of instantons ($n_+$) and anti-instantons ($n_-$) to obtain
\begin{align}
    Z(\theta, \beta) 
    = \int_{\phi(0)=0}^{\phi(\beta)=0} \mathcal{D}\phi \, e^{-S_E[\phi]} &= e^{-\omega \beta/2} \sum_{n_+, n_-=0}^\infty \frac{1}{n_+! \, n_-!} 
        \left( K_B \beta \, e^{-S_B} \right)^{n_+ + n_-} e^{i n_+ \theta} e^{-i n_- \theta} \nonumber \\\nonumber \\
    &= \exp\left[ -\frac{1}{2} \omega \beta + 2 K_B \beta \, e^{-S_B} \cos\theta \right].
    \label{eq:Z from path integral}
\end{align}
where the summation goes over all possible winding configurations. Here, $S_B$ is the bounce action, and $K_B$ arises from the functional determinant associated with small fluctuations around the instanton:
\begin{align}
    S_B &\equiv \int_0^{2\pi} d\phi \sqrt{2IV(\phi)}, \label{eq:SB}\\
    K_B &\equiv \sqrt{\frac{S_B}{2\pi}} \left| \frac{ \det(-\partial_t^2 + \omega^2) }{\det'(-\partial^2 + V''( \phi_B)/I)} \right|^{1/2}. \label{eq:KB}
\end{align}
where $\det'$ indicates that the zero mode is omitted in the determinant. 

We now show that $S_B$ and $K_B$ correspond precisely to $S$ and $K=k\omega$ defined in the semiclassical Schrödinger analysis (eqs.~\eqref{eq:S from Schroedinger eq} and \eqref{eq:K from Schroedinger eq}).  
Clearly, $S_B$ is equivalent to $S$ up to a rescaling of variables. For $K_B$, the determinant can be expressed as~\cite{Gelfand:1959nq,Coleman:1985rnk}
\begin{align}
    K_B = \sqrt{\frac{I\omega}{\pi}} A,
\end{align}
You can refer to appendix \ref{sec:KB formula} for a detailed calculation of $K_B$. Here, $A$ is a constant related to the behavior of $\phi_B(\tau)$ as discussed in~\cite{Gelfand:1959nq, Coleman:1985rnk} and characterizes the asymptotic decay of the instanton:
\begin{align}
    \frac{d\phi_B}{d\tau} \to A e^{-\omega |\tau|}. \qquad (\tau\to\pm\infty)
\end{align}
On the other hand, the Schrödinger-based quantity $K$, eq.~\eqref{eq:K from Schroedinger eq}, can be rewritten as
\begin{align}
	K = k \omega
	&\simeq \frac{\epsilon \omega}{\sqrt{\pi}} \exp\left(  \int_\epsilon^{p/2} dz' \frac{1}{\sqrt{2v(z')}}\right).
\end{align}
Introducing the inverse of the bounce, $\tau_B(\phi)$ such that $\tau_B(\phi_B(\tau)) = \tau$, we have
\begin{align}
    \int_\epsilon^{p/2} dz' \frac{1}{\sqrt{2v(z')}}
    ~=~ \omega \left[ \tau_B\left( \sqrt{\frac{1}{I\omega}}\frac{p}{2} \right) - \tau_B\left( \sqrt{\frac{1}{I\omega}}\epsilon \right) \right] 
    ~\simeq~ \frac{A}{\epsilon} \log \sqrt{I\omega}.
\end{align}
which reproduces the same prefactor as $K = \sqrt{\frac{I\omega}{\pi}} A = K_B$.
Thus we have shown that this $K_B$ is equivalent to $K$ in eq.~\eqref{eq:K from Schroedinger eq}. 

Finally, the ground-state energy and partition function derived from the path integral are
\begin{align}
    E_0 = -\lim_{\beta\to\infty} \frac{1}{\beta} \log Z(\theta, \beta) = \frac{1}{2} \omega -  2 K_B e^{-S_B} \cos\theta. \label{eq:E0 from path integral}
\end{align}
which is fully consistent with the result obtained from the semiclassical Schrödinger method (eq.~\eqref{eq:E0 from schroedinger eq}). Similarly, the topological susceptibility $\chi$ obtained from this path integral agrees with eq.~\eqref{eq:chi from schroedinger eq}.

\subsection{The partition function and topological susceptibility by ACGT method}

It is important to emphasize that the ground-state energy $E_0$ in eq.~\eqref{eq:E0 from path integral} is obtained because all possible instanton and anti-instanton configurations have been summed in $Z$ (eq.~\eqref{eq:Z from path integral}). In contrast, the approach of refs.~\cite{Ai:2020ptm,Ai:2024cnp, Ai:2025quf} considers the limit $\beta \to \infty$ before summing over all winding numbers. This subtle difference leads to different conclusions for the ground-state energy and topological susceptibility.

The partition function in eq.~\eqref{eq:Z from path integral} can be decomposed into contributions with finite winding number $\Delta$:
\begin{align}
    Z(\theta, \beta) = \sum_{\Delta = -\infty}^\infty Z_\Delta(\beta) e^{i\Delta \theta}, \label{eq:Z from Zdelta}
\end{align}
where $Z_\Delta(\beta)$ is the contribution from paths with winding number $\Delta$. Using the series representation of the modified Bessel function of the first kind,
\begin{equation}
    I_n(x) = \sum_m \frac{(2x)^{n+2m}}{[m! (m+n)!]},
\end{equation}
one finds~\cite{Ai:2020ptm, Ai:2024cnp, Khoze:2025auv} 
\begin{align}
    Z_\Delta(\beta) 
    &= \sum_{m=0}^\infty \frac{e^{-\omega \beta/2}}{|\Delta|! (|\Delta|+m)!} \left( K \beta e^{-S} \right)^{|\Delta|+2m} 
    = e^{-\omega \beta/2} I_{|\Delta|} \left( 2 K \beta e^{-S} \right).
\end{align}
We define a truncated partition function including only winding numbers $|\Delta'|\le \Delta$:
\begin{align}
    Z(\theta, \beta; \Delta) \equiv \sum_{\Delta'=-\Delta}^{\Delta} Z_{\Delta'}(\beta) e^{i \Delta' \theta}.
\end{align}
The full partition function is recovered as $\lim_{\Delta \to \infty} Z(\theta, \beta; \Delta)$. The ACGT proposal corresponds to first taking the limit $\beta \to \infty$ with fixed $\Delta$ before letting $\Delta \to \infty$.

From eq.~\eqref{eq:Z from Zdelta}, each $Z_\Delta(\beta)$ can be expressed as
\begin{align}
    Z_\Delta(\beta) = \frac{1}{2\pi} \int_0^{2\pi} d\theta \, e^{-i \Delta \theta} Z(\theta, \beta).
\end{align}
This allows rewriting the truncated partition function as
\begin{align}
    Z(\theta, \beta; \Delta) = \frac{1}{2\pi} \int_0^{2\pi} d\theta' \, f(\theta'; \theta, \Delta) \, Z(\theta', \beta), 
    \label{eq:Z with finite delta from Ztheta}
\end{align}
where
\begin{align}
    f(\theta'; \theta, \Delta) \equiv \sum_{\Delta'=-\Delta}^{\Delta} e^{i \Delta' (\theta - \theta')}.
\end{align}
Applying the dilute gas instanton approximation to eq.~\eqref{eq:Z with finite delta from Ztheta}, we obtain
\begin{align}
    Z(\theta, \beta; \Delta) 
    &= \frac{1}{2\pi} \int_0^{2\pi} d\theta' \, f(\theta'; \theta, \Delta) \, 
        \exp\Big[-\tfrac{1}{2} \omega \beta + 2 K \beta e^{-S} \cos \theta' \Big].
\end{align}
For large $\beta$ with finite $\Delta$, the integral is dominated by $\theta' \simeq 0$. Expanding around this saddle point gives
\begin{align}
    Z(\theta, \beta; \Delta) 
    &\simeq \frac{f(0; \theta, \Delta)}{2\pi} \int d\theta' \,
        \exp\Big[ -\tfrac{1}{2} \omega \beta + 2 K \beta e^{-S} \big( 1 - \tfrac{1}{2} \theta'^2 \big) 
        + \frac{f'(0; \theta, \Delta)}{f(0; \theta, \Delta)} \theta' \Big] \nonumber \\
    &= \frac{1}{2} \sqrt{\frac{\pi}{K \beta e^{-S}}} 
        \exp\Big[ -\tfrac{1}{2} \omega \beta + 2 K \beta e^{-S} + \frac{(f'(0; \theta, \Delta)/f(0; \theta, \Delta))^2}{4 K \beta e^{-S}} \Big].
\end{align}
We observe that the functional form of the prefactor $\sqrt{\pi/KT e^{-S}}$ closely resembles the result presented in ref.~\cite{Khoze:2025auv}. From the above, the effective large-$\beta$ behavior of $-(1/\beta) \log Z$ becomes
\begin{align}
    - \frac{1}{\beta} \log Z 
    \;\to\; \frac{1}{2} \omega - 2 K e^{-S} - \frac{1}{2 \beta} \log \frac{\pi}{K \beta e^{-S}}.
    \label{eq:logZ/beta a la ACGT}
\end{align}
The last term vanishes slowly as $\beta \to \infty$ but introduces a $\log \beta$ correction, which prevents a well-defined extraction of the ground-state energy. Moreover, the $\theta$-dependence disappears, leading to a vanishing topological susceptibility in this procedure.

Let us discuss the result in eq.~\eqref{eq:logZ/beta a la ACGT}. First, we note that this result does not agree with eq.~\eqref{eq:E0 from schroedinger eq}. This suggests that the ACGT procedure does not correctly reproduce the quantum mechanics of a particle on a ring with a potential in the presence of a nonzero $\theta$ parameter.
Second, the physical interpretation of eq.~\eqref{eq:Z with finite delta from Ztheta} appears to be subtle. The first two terms on the right-hand side in eq.~\eqref{eq:logZ/beta a la ACGT} coincide with the ground state energy at $\theta = 0$. Technically, this behavior can be traced to the fact that, for finite $\Delta$, $Z(\theta,T;\Delta)$ can be written in the form of eq.~\eqref{eq:Z with finite delta from Ztheta}, and that $Z(\theta',\beta)$ with $\theta'$ close to zero provides the dominant contribution in the large $T$ limit. However, eq.~\eqref{eq:Z with finite delta from Ztheta} represents a sum over partition functions $Z(\theta, \beta)$ with different values of $\theta$. Since each value of $\theta$ corresponds to a distinct superselection sector, it is not obvious how to assign a direct physical meaning to such a summation.
Furthermore, as already pointed out in ref.~\cite{Khoze:2025auv}, the expression in eq.~\eqref{eq:logZ/beta a la ACGT} contains a $\log \beta$ term, which prevents taking a well-defined limit $\beta \to \infty$. The appearance of this logarithmic term can be understood from the fact that the $\theta$ states form a continuous spectrum, so that states with $\theta$ near zero contribute dominantly to the partition function $Z$.

\section{Order of Limits, Topological Sectors, and \texorpdfstring{$\theta$}{theta}-vacua}
\label{sec:conclusion1}

In sections~\ref{sec:quantum-rotor} and \ref{sec:quantum-pendulum}, we studied simple examples of one-dimensional quantum mechanics and showed that summation over all possible configurations is required to obtain consistent results for a finite time interval $T$. We also found that imposing a restriction on the winding number leads to a mixture of different $\theta$ states. Note that a similar argument has been done in QCD \cite{Brower:2003yx}.
Since only the lowest-energy state contributes to the partition function in the limit $T \to \infty$, the two limits $T \to \infty$ and $\Delta \to \infty$ do not commute.

These observations are useful for understanding the arguments presented in the ACGT papers~\cite{Ai:2020ptm, Ai:2024vfa, Ai:2024cnp, Ai:2025quf}.  
Let us denote the state localized at $\phi \simeq 2\pi n$ by $|2\pi n\rangle$ and the shift transformation $\phi \to \phi + 2\pi$ by $G$, which is equivalent to $\exp(2\pi i p)$ used in section \ref{sec:Top_Susceptibility}.
The states $|2\pi n\rangle$, the transformation $G$, and the instantons discussed in section~\ref{sec:quantum-pendulum} correspond to the $n$ vacua, large gauge transformations, and instantons in gauge theory, respectively.
The transformation generated by $G$ is a gauge transformation in this system, because the physical region of $\phi$ is $[0,2\pi)$ and extending $\phi$ to $\mathbb{R}$ merely introduces a redundancy in the description of the physics.  
The operator $G$ commutes with the Hamiltonian $H$, and the Hilbert space decomposes into subspaces in which $G$ has eigenvalue $e^{i\theta}$.  
A superselection rule applies in each sector, and one must choose a single such subspace.
The states $|2\pi n\rangle$ themselves are not invariant under $G$, since
\begin{align}
    G |2\pi n \rangle = |2\pi(n+1)\rangle.
\end{align}
Instead, one can construct ``$\theta$ states,'' which satisfy $G|\theta\rangle = e^{i\theta}|\theta\rangle$, as
\begin{align}
    |\theta\rangle \equiv \sum_n e^{in\theta} |2\pi n\rangle.
\end{align}
This state has the lowest energy within the Hilbert subspace characterized by $G = e^{i\theta}$, and this construction is consistent with the twisted boundary condition given in eq.~\eqref{eq:twist bdry condtion}.  
We have seen that, in order to define the partition function for a given sector, one must sum over all topological sectors \textit{before} taking the limit $T \to \infty$.  
The construction of $\theta$ vacua in gauge theory proceeds in a parallel manner.
If, instead, one performs the sum over topological sectors \textit{after} taking the limit $T \to \infty$, one is led to consider the partition function $Z$ with finite $\Delta$, as discussed in sections~\ref{sec:quantum-rotor} and \ref{sec:quantum-pendulum}.  
For finite $\Delta$, the partition function $Z$ can be expressed as a linear combination of partition functions associated with different values of $\theta$, even though these sectors cannot communicate due to the superselection rule.  
In the limit $VT \to \infty$, the dominant contribution arises from the $\theta$ vacuum with $\theta \simeq 0$.  
The prefactor $(VT)^{-1/2}$ found in ref.~\cite{Khoze:2025auv} can also be understood as a consequence of the fact that the $\theta$ vacua form a continuous spectrum.

The conventional formulation of the path integral for a finite time interval $T$ is obtained by inserting a large number of completeness relations,
\begin{align}
    \langle \phi_f | e^{-HT } |\phi_i \rangle
    &= \lim_{N\to\infty}
    \left( \prod_{k'=1}^{N-1} \int_0^{2\pi} d\phi_{k'} \right)
    \left( \prod_{k''=1}^N \sum_{p_{k''}=-\infty}^\infty \right)
    \prod_{k=1}^N
    \langle \phi_k | e^{-HT  N} |p_k\rangle
    \langle p_k | \phi_{k-1}\rangle, \label{eq:path integral definition}
\end{align}
where we identify $\phi_0 = \phi_i$ and $\phi_N = \phi_f$.  
After integrating over the momentum variables $p_{k''}$, we obtain the path-integral expression, such as eq.~\eqref{eq:Z from path integral}.  
This formulation shows that one must sum over \textit{all} possible paths $\phi(\tau)$ satisfying $\phi(0) = \phi_i$ and $\phi(T) = \phi_f$.
This implies that, once $\phi = 0$ and $\phi = 2\pi$ are identified as physically equivalent points, one must sum over \textit{all} topological sectors.  
In sections~\ref{sec:quantum-rotor} and \ref{sec:quantum-pendulum}, we explicitly demonstrated that consistent results for quantum mechanics on a ring are obtained only by summing over all topological sectors.  
The correct large $T$ behavior is therefore obtained by consistently calculating at finite $T$ and then taking the limit $T \to \infty$.

\section{Conclusion} \label{sec:conclusion}
In this work we critically examined the proposal of Ai, Cruz, Garbrecht, and Tamarit that the $\theta$ dependence of gauge theories can be eliminated by a particular order-of-limits prescription in the Euclidean path integral. Using fully controlled examples in one-dimensional quantum mechanics, namely the quantum rotor and the quantum pendulum, we employed canonical quantization as a benchmark, whose results are well established and contrasted them with the path-integral formulation, which allows one to explicitly distinguish the order in which limits are taken. This framework allows us to directly test the ACGT prescription and to show that it fails to correctly reproduce the propagator, partition function, and, most importantly, the energy spectrum and topological susceptibility.

For the quantum rotor, we showed that canonical quantization and the Euclidean path-integral formulation all yield the same result for the ground-state energy and the topological susceptibility, $\chi = 1/(4\pi^2 I)$. This nonvanishing susceptibility follows directly from the $\theta$ dependence of the ground-state energy. To probe the ACGT order-of-limits prescription, we introduced a finite cutoff on the winding number as it makes the order of performing the summation over topological sectors and the infinite-time limit, explicit. We found that taking the limit $\beta \to \infty$ at fixed winding number (or within any finite subset of winding sectors) forces the susceptibility to vanish, whereas performing the sum over all winding sectors prior to the infinite-time limit reproduces the exact quantum-mechanical result. The two limits do not commute, and the physically correct order is uniquely fixed by consistency with the exact spectrum.

We also clarified that the calculation of the susceptibility in ref.~\cite{Ai:2025quf}, while correct at the level of the final result, implicitly assumes a large-gauge-invariant ground state. This assumption already enforces a sum over all topological sectors and is sufficient to determine the $\theta$ dependence of the ground-state energy independently of any path-integral considerations. Consequently, that calculation cannot be used to justify the ACGT path-integral prescription, which instead corresponds to imposing gauge-variant boundary conditions at finite spacetime volume and restoring gauge invariance only after taking the infinite-volume limit. Finally, we showed that attempts to obtain a nonzero susceptibility within a fixed topological sector rely on an unjustified interchange of the infinite-time limit with the temporal integration; when treated correctly, such contributions vanish identically. The nonzero susceptibility of the quantum rotor is therefore a genuinely global effect, originating from the unrestricted sum over winding sectors.

The quantum pendulum provides a complementary test of the ACGT prescription. In this system, the ground-state energy and its $\theta$ dependence arise from tunneling between distinct minima of the potential and are correctly reproduced only when \emph{all} instanton and anti-instanton configurations are included in the path integral. Implementing the ACGT order of limits leads to a truncated partition function that effectively mixes different $\theta$ superselection sectors. This procedure yields an expression dominated by contributions from $\theta' \simeq 0$, eliminates the physical $\theta$ dependence, and introduces a logarithmic dependence on $\beta$ that obstructs the extraction of a well-defined ground-state energy. The resulting expression disagrees with both the exact solution of the Schrödinger equation and the standard path-integral result.

The failure can be traced to the fact that truncating the winding sum amounts to averaging over distinct quantum theories labeled by $\theta$, which have no direct physical meaning when combined. As a result, the ACGT prescription does not describe the quantum mechanics of a particle on a ring with a potential in the presence of a nonzero $\theta$ parameter. The quantum pendulum thus makes explicit that the order-of-limits issue is not a technical subtlety but leads to a qualitatively incorrect theory.

Furthermore, we clarified the conceptual origin of the noncommutativity between the infinite-time limit and the sum over topological sectors by reformulating the problem in terms of gauge redundancy and superselection sectors. Using quantum mechanics on a ring as a concrete analogue of gauge theory, we showed that large gauge transformations generate a decomposition of the Hilbert space into distinct $\theta$ sectors, each characterized by a gauge-invariant ground state. Consistency of the path integral at finite time requires summing over all topological sectors before taking the $T\to\infty$ limit; reversing this order leads to a partition function that mixes distinct $\theta$ superselection sectors and is therefore unphysical. This perspective makes explicit that the ACGT prescription effectively averages over inequivalent quantum theories and explains why, in the infinite-volume limit, the result is dominated by $\theta\simeq0$ rather than describing a well-defined $\theta$ vacuum.

Taken together, our results demonstrate that the proposed elimination of $\theta$ dependence via a particular order of limits in the Euclidean path integral is incompatible with the basic structure of quantum theory. In fully solvable one-dimensional systems, we showed that restricting topological sectors or interchanging limits leads either to vanishing topological susceptibility or to ill-defined ground-state energies, in direct conflict with exact spectra and canonical quantization. The correct $\theta$ dependence emerges only as a genuinely global effect, arising from the unrestricted sum over topological sectors and the proper construction of $\theta$ vacua.
The origin of this order of limits can be directly understood from the definition of path integral given in eq.~\eqref{eq:path integral definition}, and our analysis showed that the energy spectrum and the topological susceptibility obtained from this formulation are consistent with physical properties of the system understood from gauge invariance and superselection rule.
These lessons strongly suggest that analogous prescriptions in gauge theories fail for the same reason: they do not define a single quantum theory at fixed $\theta$, but instead average over distinct ones, thereby obscuring rather than resolving the physics of the $\theta$ parameter.

The purpose of the present work is to investigate the consistency of the order-of-limits prescription proposed by ACGT~\cite{Ai:2020ptm,Ai:2024cnp,Ai:2025quf} by testing it in systems where the full quantum theory is exactly known. Our argument shows that the ACGT prescription leads inconsistent results in calculable setups. Thus, it is enough to claim that ACGT argument on non-existence of strong CP violation cannot be justified. Nevertheless, our analysis does not exclude the possibility that CP symmetry might still be conserved in strong interactions for other reasons; it only shows that such a conclusion cannot be supported by the ACGT prescription.

\section*{Acknowledgements}
This work is supported in part by JSPS KAKENHI Grant Numbers~23K03415 (RS), 24H02236 (RS), and 24H02244 (MA and RS).

\appendix
\section{Aharonov-Bohm effect for a charged particle on a ring}\label{sec:charged particle}
The Lagrangian of a particle with mass $m$ and charge $e$ is written as
\begin{align}
	L = \frac{1}{2} m (\dot x^2 + \dot y^2 + \dot z^2) - e V(\vec x) + e [A_x(\vec x) \dot x + A_y(\vec x) \dot y + A_z(\vec x) \dot z],
\end{align}
where $V(\vec x)$ and $\vec A(\vec x)$ are the electromagnetic scalar and vector potential.
If a particle is constrained to move on a ring with radius $R$ as $(x,y,z) = (R\cos\phi, R\sin\phi,0)$, then, the Lagrangian becomes 
\begin{align}
    L = \frac{1}{2} m R^2 \dot\phi^2 -e V(\phi)+ e R A_\phi(\phi) \dot\phi,
\end{align}
where $V(\phi)$ and $A_\phi(\phi) \equiv -A_x \sin\phi + A_y \cos\phi$ are the scalar and vector potential on the ring.
Let us define $\Phi$ as the magnetic flux inside the ring with radius $R$. Then, by choosing an appropriate gauge, we can take $A_\phi$ as a $\phi$-independent constant as $A_\phi = \Phi/2\pi R$. Then, by defining the moment of inertia $I=m R^2$, we can write the Lagrangian as
\begin{align}
    L = \frac{1}{2} I \dot\phi^2 -e V(\phi) + \frac{e\Phi}{2\pi}\dot\phi.
\end{align}
Then, we obtain the Hamiltonian as $H = (1/2I)\left( p_\phi - e\Phi/2\pi \right)^2 + eV(\phi)$ and the Schr\"odinger equation becomes,
\begin{align}
    \left[ \frac{1}{2I}\left( -i \frac{\partial}{\partial\phi} - \frac{e\Phi}{2\pi} \right)^2 + e V(\phi) \right] \psi(\phi) = E\psi(\phi).
\end{align}
The wave function $\psi(\phi)$ can be understood as a part of wavefunction in three-dimensional space as $\Psi(x,y,z) \sim \psi(\phi) \delta(\sqrt{x^2+y^2}-R) \delta(z)$.
Then, it is convenient to regard $\psi(\phi)$ as a function defined in $\phi \in \mathbb{R}$ with the following boundary condition:
\begin{align}
    \psi(\phi) = \psi(\phi + 2\pi).
\end{align}
It is convenient to remove the explicit flux dependence from the Hamiltonian by defining
\begin{align}
    \tilde\psi(\phi) \equiv
    \exp\!\left( - i \frac{e\Phi}{2\pi}\phi \right)\psi(\phi).
\end{align}
This corresponds to a (multi-valued) gauge transformation. The Schr\"odinger equation then takes the free-particle form
\begin{align}
    \left[ -\frac{1}{2I} \frac{\partial^2}{\partial\phi^2}
    + e V(\phi) \right] \tilde\psi(\phi)
    = E \tilde\psi(\phi),
\end{align}
while the effect of the magnetic flux is encoded in the twisted boundary condition
\begin{align}
    \tilde\psi(\phi + 2\pi)
    = \exp(- i e \Phi)\, \tilde\psi(\phi).
\end{align}
If $V(\phi)$ is independent on $\phi$ as $V(\phi) = V$, the Scrh\"odinger equation with twisted boundary conditions can be easily solved.
Then, we find 
\begin{align}
    \tilde\psi_n(\phi)=e^{i k_n \phi}, \qquad {\rm with} \quad k_n = n - \frac{e\Phi}{2\pi},
    \qquad n \in \mathbb{Z}.
\end{align}
The corresponding energy eigenvalues are
\begin{align}
    E_n(\Phi) = \frac{1}{2I}
    \left( n - \frac{e\Phi}{2\pi} \right)^2 + eV.
\end{align}
Under a shift \(\Phi \to \Phi + 2\pi/e\), the spectrum is invariant up to a relabeling \(n \to n-1\), and is therefore periodic in the flux with period \(2\pi/e\),
which is the hallmark of the Aharonov--Bohm effect.

\section{Functional determinant formula for \texorpdfstring{$K_B$}{KB}} \label{sec:KB formula}
In this appendix, we calculate $K_B$ in eq.~\eqref{eq:KB}.
First, let us define the ratio of functional determinant for a bounded function $W_i(\tau)$ as
\begin{align}
     \det ( -\partial_\tau^2 + W_i ) = \prod_n \lambda_{i,n}
\end{align}
where $\lambda_{i,n}$ is eigenvalues of $( -\partial_\tau^2 + W_i )$ acting on the space of function $f(\tau)$ such that $f(\pm T/2) = 0$.
The functional determinant satisfies the following identity 
\cite{Gelfand:1959nq, Coleman:1985rnk}:
\begin{align}
    \frac{ \det ( -\partial_\tau^2 + W_1 ) }{ \det ( -\partial_\tau^2 + W_2 ) }
    =
    \frac{\psi_{1}(T/2; 0)}{\psi_{2}(T/2; 0)},
\end{align}
Here $\psi_{i}(\tau; \lambda)$ satisfies the following differential equation and boundary conditions
\begin{align}
    (-\partial_\tau^2 + W_i - \lambda)\psi_{i}(\tau;\lambda) = 0, \label{eq:diff eq for psi0}\\
    \psi_{i}(-T/2; \lambda) = 0, \qquad
    \partial_\tau \psi_{i}(-T/2; \lambda) = 1. \label{eq:bdry psi0}
\end{align}
We assume the asymptotic behavior of $W_i(\tau)$ as
$\lim_{\tau\to\pm \infty} W_i(\tau) = \omega^2$.
Let us construct $\psi_i(\tau;0)$.
In general, eq.~\eqref{eq:diff eq for psi0} has two independent solutions. 
For $\lambda=0$, 
we can take two independent solution $x_i(\tau)$ and $y_i(\tau)$ such that their asymptotic behavior in the limit of $\tau\to\pm\infty$ are
\begin{align}
    x_i(\tau) \to a_i e^{-\omega|\tau|},\qquad
    y_i(\tau) \to \pm b_i e^{\omega|\tau|},
\end{align}
where $a_i$ and $b_i$ are some constants.
Then, for large $T$, $\psi_i(\tau; 0)$ to satisfy eq.~\eqref{eq:bdry psi0} can be constructed as
\begin{align}
    \psi_{i}(\tau; 0) = \frac{1}{2\omega} \left( \frac{e^{\omega T/2}}{a_i} x_i(\tau) + \frac{e^{-\omega T/2}}{b_i} y_i(\tau)  \right). \label{eq:psi0}
\end{align}
Then, for small $\lambda$, $\psi_{i}(\tau; \lambda)$ to satisfy eq.~\eqref{eq:bdry psi0} can be constructed from $\psi_i(\tau;0)$ and a correction term of the order of $\lambda$ as
\begin{align}
    \psi_{i}(\tau; \lambda)
    =
    \psi_{i}(\tau;0) - \frac{\lambda}{2a_i b_i \omega} \int_{-T/2}^\tau d\tau' \left[ y_i(\tau) x_i(\tau') - x_i(\tau) y_i(\tau') \right] \psi_{i}(\tau'; 0) + {\cal O}(\lambda^2).
\end{align}
Plugging eq.~\eqref{eq:psi0} into the above equation, we obtain
\begin{align}
    \psi_{i}(T/2; \lambda) = \frac{1}{\omega} - \frac{\lambda}{4\omega^2} \int_{-T/2}^{T/2} d\tau \left[ e^{\omega T} \left( \frac{x_i(\tau)}{a_i}\right)^2 - e^{-\omega T} \left( \frac{y_i(\tau)}{b_i}\right)^2 \right] + {\cal O}(\lambda^2).
\end{align}
For large $T$, we can drop the second term in the parenthesis.
We obtain the smallest eigenvalue $\lambda_{i,0}$ of $-\partial_\tau^2 + W_i(\tau)$ from $\psi_{i}(T/2; \lambda_{i,0}) = 0$ as
\begin{align}
    \lambda_{i,0} \simeq 4a_i^2 \omega e^{-\omega T} \left( \int_{-T/2}^{T/2} d\tau [x_i(\tau)]^2 \right)^{-1}.
\end{align}
Let us consider the case with $W_2(\tau) = \omega^2$.
From eq.~\eqref{eq:diff eq for psi0} and eq.~\eqref{eq:bdry psi0}, we obtain $\psi_2(\tau; 0) = (\sinh \omega T)/\omega$, and for large $T$, we obtain
\begin{align}
    \frac{\det(-\partial_\tau^2 + W_1)}{\det(-\partial_\tau^2 + \omega^2)}
    = 
    \frac{\psi_1(T/2; 0)}{(\sinh \omega T)/\omega}
    \simeq
    2 e^{-\omega T}.
\end{align}
In the limit of $T\to\infty$, 
\begin{align}
    \frac{\det'(-\partial_\tau^2 + W_1)}{\det(-\partial_\tau^2 + \omega^2)}
    = 
    \lim_{T\to\infty} \frac{1}{\lambda_{1,0}}\frac{\det(-\partial_\tau^2 + W_1)}{\det(-\partial_\tau^2 + \omega^2)}
    =
    \frac{1}{2a_1^2\omega} \int_{-T/2}^{T/2} d\tau (x_1(\tau))^2.
\end{align}
Applying this formula to eq.~\eqref{eq:KB} by replacing $a_1$,  $x_1(\tau)$, and $W_1(\tau)$ to $A$, $d\phi_B(\tau)/d\tau$, and $V(\bar\phi(\tau))/I$, we obtain
\begin{align}
    \frac{\det'(-\partial_\tau^2 + V''(\phi_B)/I)}{\det(-\partial_\tau^2 + \omega^2)}
    =
    \frac{1}{2A^2 \omega} \frac{ S_B}{I},
\end{align}
and
\begin{align}
    K_B
    \equiv \sqrt{\frac{S_B}{2\pi}}\left|\frac{\det(-\partial_\tau^2 + \omega^2)}{\det'(-\partial_\tau^2 + V''(\phi_B)/I)}\right|^{1/2}
    = \sqrt{\frac{I\omega}{\pi}} A.
\end{align}

\bibliography{ref}

@article{Callan:1976je,
    author = "Callan, Jr., Curtis G. and Dashen, R. F. and Gross, David J.",
    editor = "Taylor, J. C.",
    title = "{The Structure of the Gauge Theory Vacuum}",
    reportNumber = "COO-2220-75",
    doi = "10.1016/0370-2693(76)90277-X",
    journal = "Phys. Lett. B",
    volume = "63",
    pages = "334--340",
    year = "1976"
}

@article{Jackiw:1976pf,
    author = "Jackiw, R. and Rebbi, C.",
    editor = "Taylor, J. C.",
    title = "{Vacuum Periodicity in a Yang-Mills Quantum Theory}",
    reportNumber = "MIT-CTP-548",
    doi = "10.1103/PhysRevLett.37.172",
    journal = "Phys. Rev. Lett.",
    volume = "37",
    pages = "172--175",
    year = "1976"
}

@article{Ai:2020ptm,
    author = {Ai, Wen-Yuan and Cruz, Juan S. and Garbrecht, Bj{\"o}rn and Tamarit, Carlos},
    title = "{Consequences of the order of the limit of infinite spacetime volume and the sum over topological sectors for CP violation in the strong interactions}",
    eprint = "2001.07152",
    archivePrefix = "arXiv",
    primaryClass = "hep-th",
    reportNumber = "TUM-HEP-1249/20, CP3-20-02",
    doi = "10.1016/j.physletb.2021.136616",
    journal = "Phys. Lett. B",
    volume = "822",
    pages = "136616",
    year = "2021"
}

@article{Ai:2024cnp,
    author = "Ai, Wen-Yuan and Garbrecht, Bjorn and Tamarit, Carlos",
    title = "{CP Conservation in the Strong Interactions}",
    eprint = "2404.16026",
    archivePrefix = "arXiv",
    primaryClass = "hep-ph",
    reportNumber = "KCL-PH-TH/2023-51, MITP-24-043, TUM-HEP-1508/24",
    doi = "10.3390/universe10050189",
    journal = "Universe",
    volume = "10",
    number = "5",
    pages = "189",
    year = "2024"
}

@article{Ai:2025quf,
    author = {Ai, Wen-Yuan and Garbrecht, Bj{\"o}rn and Tamarit, Carlos},
    title = "{Reply to ''Clearing up the Strong $CP$ problem''}",
    eprint = "2511.04216",
    archivePrefix = "arXiv",
    primaryClass = "hep-ph",
    reportNumber = "MITP-25-070, TUM-HEP-1576/25",
    month = "11",
    year = "2025"
}

@article{Benabou:2025viy,
    author = "Benabou, Joshua N. and Hook, Anson and Manzari, Claudio Andrea and Murayama, Hitoshi and Safdi, Benjamin R.",
    title = "{Clearing up the Strong $CP$ problem}",
    eprint = "2510.18951",
    archivePrefix = "arXiv",
    primaryClass = "hep-ph",
    month = "10",
    year = "2025"
}

@article{Khoze:2025auv,
    author = "Khoze, Valentin V.",
    title = "{A note on instantons, $\theta$-dependence and strong CP}",
    eprint = "2512.06827",
    archivePrefix = "arXiv",
    primaryClass = "hep-ph",
    reportNumber = "IPPP-25-XXX",
    month = "12",
    year = "2025"
}

@article{Albandea:2024fui,
    author = "Albandea, David and Catumba, Guilherme and Ramos, Alberto",
    title = "{Strong CP problem in the quantum rotor}",
    eprint = "2402.17518",
    archivePrefix = "arXiv",
    primaryClass = "hep-lat",
    doi = "10.1103/PhysRevD.110.094512",
    journal = "Phys. Rev. D",
    volume = "110",
    number = "9",
    pages = "094512",
    year = "2024"
}

@article{Nakamura:2021meh,
    author = "Nakamura, Y. and Schierholz, G.",
    title = "{The strong CP problem solved by itself due to long-distance vacuum effects}",
    eprint = "2106.11369",
    archivePrefix = "arXiv",
    primaryClass = "hep-ph",
    reportNumber = "DESY 21-078, DESY-21-078",
    doi = "10.1016/j.nuclphysb.2022.116063",
    journal = "Nucl. Phys. B",
    volume = "986",
    pages = "116063",
    year = "2023"
}

@article{Schierholz:2024var,
    author = "Schierholz, Gerrit",
    title = "{Absence of strong CP violation}",
    eprint = "2403.13508",
    archivePrefix = "arXiv",
    primaryClass = "hep-ph",
    reportNumber = "DESY-24-038",
    doi = "10.1088/1361-6471/adc31d",
    journal = "J. Phys. G",
    volume = "52",
    number = "4",
    pages = "04LT01",
    year = "2025"
}

@article{Schierholz:2025tns,
    author = "Schierholz, Gerrit",
    title = "{Absence of CP Violation in the Strong Interaction: Vacuum thwarts Axion}",
    eprint = "2502.04092",
    archivePrefix = "arXiv",
    primaryClass = "hep-lat",
    reportNumber = "DESY-25-21",
    doi = "10.22323/1.466.0398",
    journal = "PoS",
    volume = "LATTICE2024",
    pages = "398",
    year = "2025"
}

@article{Bhattacharya:2025qsk,
    author = "Bhattacharya, Tanmoy",
    title = "{Comment on the claim of physical irrelevance of the topological term}",
    eprint = "2512.10127",
    archivePrefix = "arXiv",
    primaryClass = "hep-lat",
    reportNumber = "LA-UR-22-29996",
    month = "12",
    year = "2025"
}

@article{Yamanaka:2022vdt,
    author = "Yamanaka, Nodoka",
    title = "{Unobservability of topological charge in nonabelian gauge theory}",
    eprint = "2212.10994",
    archivePrefix = "arXiv",
    primaryClass = "hep-th",
    month = "12",
    year = "2022"
}

@article{Yamanaka:2022bfj,
    author = "Yamanaka, Nodoka",
    title = "{Unobservability of the topological charge in nonabelian gauge theory: Ward-Takahashi identity and phenomenological aspects}",
    eprint = "2212.11820",
    archivePrefix = "arXiv",
    primaryClass = "hep-ph",
    month = "12",
    year = "2022"
}

@article{Yamanaka:2024nzn,
    author = "Yamanaka, Nodoka",
    title = "{Sketch of the resolution of the axial U(1) problem without chiral anomaly}",
    eprint = "2411.02792",
    archivePrefix = "arXiv",
    primaryClass = "hep-ph",
    month = "11",
    year = "2024"
}

@article{Shifman:1979if,
    author = "Shifman, Mikhail A. and Vainshtein, A. I. and Zakharov, Valentin I.",
    title = "{Can Confinement Ensure Natural CP Invariance of Strong Interactions?}",
    reportNumber = "ITEP-64-1979",
    doi = "10.1016/0550-3213(80)90209-6",
    journal = "Nucl. Phys. B",
    volume = "166",
    pages = "493--506",
    year = "1980"
}

@article{Gabadadze:2002ff,
    author = "Gabadadze, Gregory and Shifman, M.",
    editor = "Olive, K. A. and Shifman, M. A. and Voloshin, M. B.",
    title = "{QCD vacuum and axions: What's happening?}",
    eprint = "hep-ph/0206123",
    archivePrefix = "arXiv",
    reportNumber = "TPI-MINN-02-19, UMN-TH-2104-02",
    doi = "10.1142/S0217751X02011357",
    journal = "Int. J. Mod. Phys. A",
    volume = "17",
    pages = "3689--3728",
    year = "2002"
}

@article{Witten:1979vv,
    author = "Witten, Edward",
    title = "{Current Algebra Theorems for the U(1) Goldstone Boson}",
    reportNumber = "HUTP-79/A014",
    doi = "10.1016/0550-3213(79)90031-2",
    journal = "Nucl. Phys. B",
    volume = "156",
    pages = "269--283",
    year = "1979"
}

@article{Veneziano:1979ec,
    author = "Veneziano, G.",
    title = "{U(1) Without Instantons}",
    reportNumber = "CERN-TH-2651",
    doi = "10.1016/0550-3213(79)90332-8",
    journal = "Nucl. Phys. B",
    volume = "159",
    pages = "213--224",
    year = "1979"
}

@article{Pich:1991fq,
    author = "Pich, Antonio and de Rafael, Eduardo",
    title = "{Strong CP violation in an effective chiral Lagrangian approach}",
    reportNumber = "CERN-TH-6071-91",
    doi = "10.1016/0550-3213(91)90019-T",
    journal = "Nucl. Phys. B",
    volume = "367",
    pages = "313--333",
    year = "1991"
}

@article{Crewther:1979pi,
    author = "Crewther, R. J. and Di Vecchia, P. and Veneziano, G. and Witten, Edward",
    title = "{Chiral Estimate of the Electric Dipole Moment of the Neutron in Quantum Chromodynamics}",
    reportNumber = "CERN-TH-2735",
    doi = "10.1016/0370-2693(79)90128-X",
    journal = "Phys. Lett. B",
    volume = "88",
    pages = "123",
    year = "1979",
    note = "[Erratum: Phys.Lett.B 91, 487 (1980)]"
}

@article{Dine:2016sgq,
    author = "Dine, Michael and Draper, Patrick and Stephenson-Haskins, Laurel and Xu, Di",
    title = "{$\theta$ and the $\eta^\prime$ in Large $N$ Supersymmetric QCD}",
    eprint = "1612.05770",
    archivePrefix = "arXiv",
    primaryClass = "hep-th",
    doi = "10.1007/JHEP05(2017)122",
    journal = "JHEP",
    volume = "05",
    pages = "122",
    year = "2017"
}

@article{Csaki:2023yas,
    author = "Cs{\'a}ki, Csaba and Tito D'Agnolo, Raffaele and Gupta, Rick S. and Kuflik, Eric and Roy, Tuhin S. and Ruhdorfer, Maximilian",
    title = "{On the dynamical origin of the {\ensuremath{\eta'}} potential and the axion mass}",
    eprint = "2307.04809",
    archivePrefix = "arXiv",
    primaryClass = "hep-ph",
    doi = "10.1007/JHEP10(2023)139",
    journal = "JHEP",
    volume = "10",
    pages = "139",
    year = "2023"
}

@phdthesis{LeinoThesis,
    author = "Leinno, Markku",
    title = "{Finite-Temperature Quantum Statistics of a Few
Confined Electrons and Atoms - Path-Integral Approach}",
    school = "Tampere
University of Technology",
    year = "2007"
}

@article{Abel:2020pzs,
    author = "Abel, C. and others",
    title = "{Measurement of the Permanent Electric Dipole Moment of the Neutron}",
    eprint = "2001.11966",
    archivePrefix = "arXiv",
    primaryClass = "hep-ex",
    doi = "10.1103/PhysRevLett.124.081803",
    journal = "Phys. Rev. Lett.",
    volume = "124",
    number = "8",
    pages = "081803",
    year = "2020"
}

@article{Chupp:2017rkp,
    author = "Chupp, Timothy and Fierlinger, Peter and Ramsey-Musolf, Michael and Singh, Jaideep",
    title = "{Electric dipole moments of atoms, molecules, nuclei, and particles}",
    eprint = "1710.02504",
    archivePrefix = "arXiv",
    primaryClass = "physics.atom-ph",
    reportNumber = "ACFI-T17-18",
    doi = "10.1103/RevModPhys.91.015001",
    journal = "Rev. Mod. Phys.",
    volume = "91",
    number = "1",
    pages = "015001",
    year = "2019"
}

@article{Weinberg:1975ui,
    author = "Weinberg, Steven",
    title = "{The U(1) Problem}",
    reportNumber = "Print-75-0266 (HARVARD)",
    doi = "10.1103/PhysRevD.11.3583",
    journal = "Phys. Rev. D",
    volume = "11",
    pages = "3583--3593",
    year = "1975"
}

@book{Coleman:1985rnk,
    author = "Coleman, Sidney",
    title = "{Aspects of Symmetry}: {Selected Erice Lectures}",
    doi = "10.1017/CBO9780511565045",
    isbn = "978-0-521-31827-3",
    publisher = "Cambridge University Press",
    address = "Cambridge, U.K.",
    year = "1985"
}

@book{Rubakov:2002fi,
    author = "Rubakov, V. A.",
    title = "{Classical theory of gauge fields}",
    isbn = "978-0-691-05927-3, 978-0-691-05927-3",
    publisher = "Princeton University Press",
    address = "Princeton, New Jersey",
    month = "5",
    year = "2002",
    doi = "10.1515/9781400825097"
}

@article{Bachas:2016ffl,
    author = "Bachas, Constantin and Tomaras, Theodore",
    title = "{Band Structure in Yang-Mills Theories}",
    eprint = "1603.08749",
    archivePrefix = "arXiv",
    primaryClass = "hep-th",
    reportNumber = "LPTENS-16-01, ITCP-IPP-2016-03",
    doi = "10.1007/JHEP05(2016)143",
    journal = "JHEP",
    volume = "05",
    pages = "143",
    year = "2016"
}

@book{Schulman:1981vu,
    author = "Schulman, L. s.",
    title = "{TECHNIQUES AND APPLICATIONS OF PATH INTEGRATION}",
    year = "1981",
    isbn={9780471764502},
    lccn={80019129},
    url={https://books.google.co.jp/books?id=zRJRAAAAMAAJ},
    publisher={Wiley}
}

@article{Gelfand:1959nq,
    author = "Gelfand, I. M. and Yaglom, A. M.",
    title = "{Integration in functional spaces and it applications in quantum physics}",
    doi = "10.1063/1.1703636",
    journal = "J. Math. Phys.",
    volume = "1",
    pages = "48",
    year = "1960"
}

@article{Belavin:1975fg,
    author = "Belavin, A. A. and Polyakov, Alexander M. and Schwartz, A. S. and Tyupkin, Yu. S.",
    editor = "Taylor, J. C.",
    title = "{Pseudoparticle Solutions of the Yang-Mills Equations}",
    doi = "10.1016/0370-2693(75)90163-X",
    journal = "Phys. Lett. B",
    volume = "59",
    pages = "85--87",
    year = "1975"
}

@article{tHooft:1976snw,
    author = "'t Hooft, Gerard",
    editor = "Shifman, Mikhail A.",
    title = "{Computation of the Quantum Effects Due to a Four-Dimensional Pseudoparticle}",
    reportNumber = "PRINT-76-0551 (HARVARD)",
    doi = "10.1103/PhysRevD.14.3432",
    journal = "Phys. Rev. D",
    volume = "14",
    pages = "3432--3450",
    year = "1976",
    note = "[Erratum: Phys.Rev.D 18, 2199 (1978)]"
}

@article{tHooft:1976rip,
    author = "'t Hooft, Gerard",
    editor = "Shifman, Mikhail A.",
    title = "{Symmetry Breaking Through Bell-Jackiw Anomalies}",
    reportNumber = "PRINT-76-0254 (HARVARD)",
    doi = "10.1103/PhysRevLett.37.8",
    journal = "Phys. Rev. Lett.",
    volume = "37",
    pages = "8--11",
    year = "1976"
}

@article{Pospelov:1999ha,
    author = "Pospelov, Maxim and Ritz, Adam",
    title = "{Theta induced electric dipole moment of the neutron via QCD sum rules}",
    eprint = "hep-ph/9904483",
    archivePrefix = "arXiv",
    reportNumber = "TPI-MINN-99-24, UMN-TH-1761-99",
    doi = "10.1103/PhysRevLett.83.2526",
    journal = "Phys. Rev. Lett.",
    volume = "83",
    pages = "2526--2529",
    year = "1999"
}

@article{Pospelov:1999mv,
    author = "Pospelov, Maxim and Ritz, Adam",
    title = "{Theta vacua, QCD sum rules, and the neutron electric dipole moment}",
    eprint = "hep-ph/9908508",
    archivePrefix = "arXiv",
    reportNumber = "TPI-MINN-99-34, UMN-TH-1808-99",
    doi = "10.1016/S0550-3213(99)00817-2",
    journal = "Nucl. Phys. B",
    volume = "573",
    pages = "177--200",
    year = "2000"
}

@article{Pospelov:2005pr,
    author = "Pospelov, Maxim and Ritz, Adam",
    title = "{Electric dipole moments as probes of new physics}",
    eprint = "hep-ph/0504231",
    archivePrefix = "arXiv",
    doi = "10.1016/j.aop.2005.04.002",
    journal = "Annals Phys.",
    volume = "318",
    pages = "119--169",
    year = "2005"
}

@article{Hisano:2012sc,
    author = "Hisano, Junji and Lee, Jeong Yong and Nagata, Natsumi and Shimizu, Yasuhiro",
    title = "{Reevaluation of Neutron Electric Dipole Moment with QCD Sum Rules}",
    eprint = "1204.2653",
    archivePrefix = "arXiv",
    primaryClass = "hep-ph",
    reportNumber = "IPMU-12-0065, TU-903",
    doi = "10.1103/PhysRevD.85.114044",
    journal = "Phys. Rev. D",
    volume = "85",
    pages = "114044",
    year = "2012"
}

@article{Choi:1990cn,
    author = "Choi, Kiwoon and Hong, Joo-yoo",
    title = "{Electron electric dipole moment and Theta (QCD)}",
    reportNumber = "SNUTP-90-32",
    doi = "10.1016/0370-2693(91)90838-H",
    journal = "Phys. Lett. B",
    volume = "259",
    pages = "340--344",
    year = "1991"
}

@article{Ghosh:2017uqq,
    author = "Ghosh, Diptimoy and Sato, Ryosuke",
    title = "{Lepton Electric Dipole Moment and Strong CP Violation}",
    eprint = "1709.05866",
    archivePrefix = "arXiv",
    primaryClass = "hep-ph",
    doi = "10.1016/j.physletb.2017.12.052",
    journal = "Phys. Lett. B",
    volume = "777",
    pages = "335--339",
    year = "2018"
}

@article{Ema:2024vfn,
    author = "Ema, Yohei and Gao, Ting and Pospelov, Maxim and Ritz, Adam",
    title = "{Chiral properties of the nucleon interpolating current and {\ensuremath{\theta}}-dependent observables}",
    eprint = "2405.08856",
    archivePrefix = "arXiv",
    primaryClass = "hep-ph",
    reportNumber = "UMN-TH-4319/24, FTPI-MINN-24-10",
    doi = "10.1103/PhysRevD.110.034028",
    journal = "Phys. Rev. D",
    volume = "110",
    number = "3",
    pages = "034028",
    year = "2024"
}

@article{Flambaum:2019ejc,
    author = "Flambaum, V. V. and Pospelov, M. and Ritz, A. and Stadnik, Y. V.",
    title = "{Sensitivity of EDM experiments in paramagnetic atoms and molecules to hadronic CP violation}",
    eprint = "1912.13129",
    archivePrefix = "arXiv",
    primaryClass = "hep-ph",
    doi = "10.1103/PhysRevD.102.035001",
    journal = "Phys. Rev. D",
    volume = "102",
    number = "3",
    pages = "035001",
    year = "2020"
}

@article{Pospelov:2025vzj,
    author = "Pospelov, Maxim and Ritz, Adam",
    title = "{Electric Dipole Moments and New Physics}",
    eprint = "2509.23531",
    archivePrefix = "arXiv",
    primaryClass = "hep-ph",
    month = "9",
    year = "2025"
}

@article{Strocchi:2024tis,
    author = "Strocchi, F.",
    title = "{The strong CP problem revisited and solved by the gauge group topology}",
    eprint = "2404.19400",
    archivePrefix = "arXiv",
    primaryClass = "hep-th",
    month = "4",
    year = "2024"
}

@article{Dragos:2019oxn,
    author = "Dragos, Jack and Luu, Thomas and Shindler, Andrea and de Vries, Jordy and Yousif, Ahmed",
    title = "{Confirming the Existence of the strong CP Problem in Lattice QCD with the Gradient Flow}",
    eprint = "1902.03254",
    archivePrefix = "arXiv",
    primaryClass = "hep-lat",
    doi = "10.1103/PhysRevC.103.015202",
    journal = "Phys. Rev. C",
    volume = "103",
    number = "1",
    pages = "015202",
    year = "2021"
}

@article{Alexandrou:2020mds,
    author = "Alexandrou, C. and Athenodorou, A. and Hadjiyiannakou, K. and Todaro, A.",
    title = "{Neutron electric dipole moment using lattice QCD simulations at the physical point}",
    eprint = "2011.01084",
    archivePrefix = "arXiv",
    primaryClass = "hep-lat",
    doi = "10.1103/PhysRevD.103.054501",
    journal = "Phys. Rev. D",
    volume = "103",
    number = "5",
    pages = "054501",
    year = "2021"
}

@article{Bhattacharya:2021lol,
    author = "Bhattacharya, Tanmoy and Cirigliano, Vincenzo and Gupta, Rajan and Mereghetti, Emanuele and Yoon, Boram",
    title = "{Contribution of the QCD $\Theta$-term to the nucleon electric dipole moment}",
    eprint = "2101.07230",
    archivePrefix = "arXiv",
    primaryClass = "hep-lat",
    reportNumber = "LA-UR-20-30515",
    doi = "10.1103/PhysRevD.103.114507",
    journal = "Phys. Rev. D",
    volume = "103",
    number = "11",
    pages = "114507",
    year = "2021"
}

@article{Liang:2023jfj,
    author = "Liang, Jian and Alexandru, Andrei and Draper, Terrence and Liu, Keh-Fei and Wang, Bigeng and Wang, Gen and Yang, Yi-Bo",
    collaboration = "{\ensuremath{\chi}}QCD",
    title = "{Nucleon electric dipole moment from the {\ensuremath{\theta}} term with lattice chiral fermions}",
    eprint = "2301.04331",
    archivePrefix = "arXiv",
    primaryClass = "hep-lat",
    doi = "10.1103/PhysRevD.108.094512",
    journal = "Phys. Rev. D",
    volume = "108",
    number = "9",
    pages = "094512",
    year = "2023"
}

@article{He:2023gwp,
    author = "He, Fangcheng and Abramczyk, Michael and Blum, Tom and Izubuchi, Taku and Ohki, Hiroshi and Syritsyn, Sergey",
    title = "{The calculations of Nucleon Electric Dipole Moment using background field on Lattice QCD}",
    eprint = "2311.06106",
    archivePrefix = "arXiv",
    primaryClass = "hep-lat",
    doi = "10.22323/1.453.0336",
    journal = "PoS",
    volume = "LATTICE2023",
    pages = "336",
    year = "2024"
}

@article{Liu:2024kqy,
    author = "Liu, Keh-Fei",
    title = "{Lattice QCD and the Neutron Electric Dipole Moment}",
    eprint = "2411.15198",
    archivePrefix = "arXiv",
    primaryClass = "hep-lat",
    doi = "10.1146/annurev-nucl-121423-100927",
    journal = "Ann. Rev. Nucl. Part. Sci.",
    volume = "75",
    number = "1",
    pages = "377--397",
    year = "2025"
}

@article{Cheng:1987gp,
    author = "Cheng, Hai-Yang",
    title = "{The Strong CP Problem Revisited}",
    reportNumber = "IUHET-125-REV, IUHET-125",
    doi = "10.1016/0370-1573(88)90135-4",
    journal = "Phys. Rept.",
    volume = "158",
    pages = "1",
    year = "1988"
}

@article{KLOE-2:2020ydi,
    author = "Babusci, D. and others",
    collaboration = "KLOE-2",
    title = "{Upper limit on the $\eta\to\pi^{+}\pi^{-}$ branching fraction with the KLOE experiment}",
    eprint = "2006.14710",
    archivePrefix = "arXiv",
    primaryClass = "hep-ex",
    doi = "10.1007/JHEP10(2020)047",
    journal = "JHEP",
    volume = "10",
    pages = "047",
    year = "2020"
}

@article{Roussy:2022cmp,
    author = "Roussy, Tanya S. and others",
    title = "{An improved bound on the electron{\textquoteright}s electric dipole moment}",
    eprint = "2212.11841",
    archivePrefix = "arXiv",
    primaryClass = "physics.atom-ph",
    doi = "10.1126/science.adg4084",
    journal = "Science",
    volume = "381",
    number = "6653",
    pages = "adg4084",
    year = "2023"
}

@article{Brower:2003yx,
    author = "Brower, R. and Chandrasekharan, S. and Negele, John W. and Wiese, U. J.",
    title = "{QCD at fixed topology}",
    eprint = "hep-lat/0302005",
    archivePrefix = "arXiv",
    reportNumber = "DUKE-TH-02-229, MIT-CTP-3201",
    doi = "10.1016/S0370-2693(03)00369-1",
    journal = "Phys. Lett. B",
    volume = "560",
    pages = "64--74",
    year = "2003"
}

@article{Gamboa:2025hxa,
    author = "Gamboa, Jorge and Arellano, Natalia A. Tapia",
    title = "{Strong CP as an Infrared Holonomy: The $\theta$ Vacuum and Dressing in Yang-Mills Theory}",
    eprint = "2512.24480",
    archivePrefix = "arXiv",
    primaryClass = "hep-th",
    month = "12",
    year = "2025"
}

@inproceedings{Ringwald:2026apz,
    author = "Ringwald, Andreas",
    title = "{CP, or not CP, that is the question...}",
    booktitle = "{3rd General Meeting of the COST Action: Cosmic WISPers (CA21106)}",
    eprint = "2601.04718",
    archivePrefix = "arXiv",
    primaryClass = "hep-ph",
    reportNumber = "DESY-26-001",
    month = "1",
    year = "2026"
}

@article{Kleinert:2004ev,
    author = "Kleinert, H.",
    title = "{Path Integrals in Quantum Mechanics, Statistics, Polymer Physics, and Financial Markets}",
    year = "2004"
}

@article{Luscher:2004fu,
    author = "Luscher, Martin",
    title = "{Topological effects in QCD and the problem of short distance singularities}",
    eprint = "hep-th/0404034",
    archivePrefix = "arXiv",
    reportNumber = "CERN-PH-TH-2004-062",
    doi = "10.1016/j.physletb.2004.04.076",
    journal = "Phys. Lett. B",
    volume = "593",
    pages = "296--301",
    year = "2004"
}

@article{Luscher:2010ik,
    author = "Luscher, Martin and Palombi, Filippo",
    title = "{Universality of the topological susceptibility in the SU(3) gauge theory}",
    eprint = "1008.0732",
    archivePrefix = "arXiv",
    primaryClass = "hep-lat",
    reportNumber = "CERN-PH-TH-2010-173",
    doi = "10.1007/JHEP09(2010)110",
    journal = "JHEP",
    volume = "09",
    pages = "110",
    year = "2010"
}

@article{Ai:2024vfa,
    author = "Ai, Wen-Yuan and Garbrecht, Bjorn and Tamarit, Carlos",
    title = "{The QCD theta-parameter in canonical quantization}",
    eprint = "2403.00747",
    archivePrefix = "arXiv",
    primaryClass = "hep-th",
    reportNumber = "KCL-PH-TH/2024-13, TUM-HEP-1499/24, MITP-24-031",
    month = "3",
    year = "2024"
}
\bibliographystyle{JHEP}
\end{document}